**Who Are Tweeting About Academic Publications? A Systematic Review and Meta-Analysis of Altmetric Studies**

Ashraf Maleki[1] (https://orcid.org/0000-0002-8223-4833) and Kim Holmberg[1] (https://orcid.org/0000-0002-4185-9298)
[1]Department of Social Research, University of Turku, Turku, Finland

Corresponding Author:
Ashraf Maleki
*ashraf.maleki@utu.fi*

**Keywords:** meta-analysis, systematic review, altmetrics, X/Twitter, user categorization, scholarly communications

# 1 Abstract

Understanding who shares academic publications on Twitter is critical to interpreting altmetrics as signals of scholarly or societal impact. Prior studies have used diverse and often incompatible user classification schemes, making synthesis difficult. This study presents a systematic review and meta-analysis of 23 empirical studies (covering 79,014 Twitter users, over 20 million tweets, and more than 5 million tweeted publications) to estimate category-specific engagement across three metrics: user counts, tweets, and tweeted publications. We developed a harmonized categorization scheme encompassing 11 user types and applied both Random Effects Models (REM) and Beta-Binomial Hierarchical Models (BBHM) to estimate proportions, account for study-level variation, and model uncertainty. Across all indicators, individual users were the most active, comprising 66% of users, 55% of tweets, and 50% of tweeted publications. BBHM further enabled in-category vs. out-of-category comparisons and revealed engagement differences not detected by REM. T-tests on study-level means confirmed significant differences between academic individuals and other user types. Despite methodological heterogeneity, results consistently show that academic and non-academic individuals statistically equally dominate Twitter engagement with scholarly content. Our findings support the need for standardized user classification schemes and demonstrate the value of Bayesian modeling for synthesizing altmetric data in study variation and sparsity.

# 2 Introduction

Investigating the online mentions of research outputs, or the altmetrics of research outputs, can reveal where, how, and by whom research has been shared, disseminated, and discussed on various online platforms. Such information could lead to an increased understanding of how research may have had some type of impact or influence on wider audiences beyond academia, and with that, how altmetrics could be used for impact assessment (Holmberg, 2015). But this requires that we understand who shares, disseminates, or discusses scientific documents online and for what reason. Twitter, or X as the service is currently called, is one of the main sources of altmetrics and several earlier studies have focused on investigating the people who share scientific articles on Twitter. Because this study investigates earlier studies that were conducted before the namechange, we will keep referring to the service as Twitter. To understand whether Twitter can reflect a wider societal interest or attention towards scientific content, it is important to understand the diverse roles, motivation and behaviours of Twitter users who share and interact with scientific articles on the platform. Without user characterization, altmetrics studies risk missing out on valuable insights into the impact and reach of scholarly content on social media platforms. Because different user types may interact with content in distinct ways, without a Twitter user categorization, it would be difficult to interpret the engagement metrics. Despite Twitter's open structure, categorizing its users based on their occupation, expertise or motivation remains challenging. While some studies have identified scientists on Twitter (Schmitt & Jaschke, 2017) and their professional use of the platform (Álvarez-Bornstein & Montesi, 2019), others highlight the mix between personal and professional engagement (Bowman, 2015; Toupin, Millerand & Larivière, 2019). This variability in user classification approaches complicates cross-study comparisons and generalization of findings, potentially leading to inconsistent conclusions about who actually drives research dissemination on Twitter. Without clear and standardized user categories, altmetrics studies struggle to accurately contextualize engagement patterns, leading to difficulties in distinguishing between academic vs. non-academic attention and scholarly vs. social impact.

While numerous classification techniques have been used to explore Twitter users engaging with research, a systematic, meta-analytical synthesis of these studies is lacking. Previous studies have reviewed and condensed research findings on various topics, such as the utilization of Twitter by medical journals (Erskine & Hendricks, 2021), the association between altmetrics and citations (e.g., Erdt, Nagarajan et al., 2016; Bornmann, 2015), and the comparison of altmetric data providers (Ortega, 2020). However, no study has systematically consolidated findings regarding who tweets about scientific content and to what extent. This meta-analysis aims to bridge this gap by quantitatively synthesizing findings from prior user classification studies, and providing a clearer picture of the composition and behaviors of Twitter users engaging with academic publications.

Meta-analysis methods traditionally rely on frequentist random-effects models, but such methods have limitations when study-level variability is high or when the number of studies is small. These characteristics are particularly relevant to meta-analysis of Twitter user categorization research, where the number of studies is limited, yet many studies analyze millions of tweets. Bayesian hierarchical models (BHM) have become increasingly popular for meta-analysis of social science and biomedical data, where heterogeneity across studies and small study counts pose challenges (Schmid, 2001; Gelman, Carlin et al., 2014). A Bayesian hierarchical model is a probabilistic framework that estimates group-level effects by pooling information across studies while accounting for uncertainty and variability at multiple levels of analysis, therefore offering several advantages over frequentist approaches, including the ability to incorporate weakly-informed prior as binary variables (Negrín-Hernández, Martel-Escobar, & Vázquez-Polo, 2021) and greater flexibility in modeling study-level differences. Hierarchical Bayesian models have been applied in capturing intensity and continuity of scholars' productivity over time (Mezzetti & Negri, 2024), publication bias corrections (Schmidt & Hunter, 2015) which demonstrate their robustness in meta-analytical settings.

The Beta-Binomial Hierarchical Model (BBHM) is particularly useful in studies where success and failure events are being measured. This approach has been widely applied meta-analysis method for few studies involving rare events (Günhan, Röver, & Friede, 2020; Jansen & Holling, 2023), and identifying individual differences in population studies (Doebler, Frick, & Doebler, 2024). The Beta-Binomial structure is advantageous because it naturally models count data and gives success rate in proportion; accounts for study-level variability, ensuring that studies with different sample sizes contribute proportionally to the meta-analysis; and more importantly allows for robust inference even when the number of studies is small despite large datasets in each study, a common limitation faced in meta-analysis of altmetrics research. This approach enables us to estimate the likelihood of previous studies identifying different user types, by better estimation of uncertainty and credible intervals. In Bayesian models, full posterior distributions is provided rather than relying on point estimates and confidence intervals, which enhances the understanding of engagement patterns. Given the diverse nature of Twitter studies, Bayesian methods allow for the inclusion of weak evidence, reducing the risk of bias.

# 3 Background

Earlier studies have had different objectives for their user categorizations, leading to different approaches. Profile descriptions of Twitter users are often the basis for the coding, resulting in category dependence on the information that users share about themselves. Many studies (13) have utilized two or three coders to enhance inter-coder reliability (Yu, Xiao et al., 2019; Zhou & Na, 2019), but there are also studies that have tried to analyse broader patterns, adopting automated or semi-automated user categorization methods on larger datasets. Precision and accuracy of automatic categorizations based on keywords in Twitter profiles have been shown to have some limitations. Barthel, Tönnies et al. (2015) for instance, used an automated approach to detect scientists among a group of tweeters, reaching an accuracy of 88%, while another more recent study showed that over 95% precision can be obtained by matching the names of 500,000 scientists on ORCID and Twitter usernames on Crossref Event data (Mongeon, Bowman, & Costas, 2022). Manual coding of studies originates in researchers' interest to more accurately indicate diversity of user categories across either their individual and organizational association (Orduña-Malea & Costas, 2021) or various professions (de Melo Maricato & de Castro Manso, 2022). Many studies have been interested in bringing the focus to only particular users types that are actively involved in sharing scientific publications, such as academics (Zhou & Na, 2019), health practitioners such as surgeons (Coret, Rok et al., 2020), publishers and journal accounts (Ortega, 2017) and bots (Haustein, Bowman et al., 2016) or a combination of these (Haustein, 2019; Ye, Na, & Oh, 2022). Other data from profile descriptions, such as occupation, gender (e.g., Tsou, Bowman et al., 2015; Ke, Ahn, & Sugimoto, 2017), and geographic location (Dudek, Bowman, & Costas, 2018; Haustein, 2019; Yu, Xiao et al., 2019) have also been used for user categorization. Some studies have used a binary categorization, such as *bot* and *non-bot* (Haustein, Bowman et al., 2016; Haustein, Tsou et al., 2016; Didegah, Mejlgaard, & Sørensen, 2018; Yu, Xiao et al, 2019); *academic* and *non-academic* (Ke, Ahn, & Sugimoto, 2017; Htoo & Jin-Cheon, 2017; Didegah, Mejlgaard & Sørensen, 2018; Zhou & Na, 2019); *male* or *female* (Orduña-Malea & Costas, 2021); and *expert* or *non-expert* (Pandian, Devi et al., 2019). Other studies have used three categories, such as *individual*, *organization*, and *the rest*, which have included unknown or unidentifiable accounts (Haustein et al., 2016b), or *professional*, *non-professional*, and *composite*, where the categorization was based on the professionalism expressed in the profile description (Yu et al., 2019). A few studies have taken a more detailed approach categorizing Twitter users into more numerous categories in order to broaden the perspective of the types of users involved in sharing scientific articles, such as the nine categories used by Haustein (2020) and the 27 categories used by Vainio & Holmberg (2017). On the other hand, many studies have used the categorization provided by the altmetric data provider *Altmetric.com*, that identifies only some specific types of users, such as *scientists*, *practitioners*, and *science communicators*, everything else being categorized as *members of the public*.

Overall, earlier studies have shown that users who tweet scientific content can come from a range of backgrounds and include both scholars and members of the public (Haustein, Costas, and Larivière, 2015), but some studies have shown that academics may be overrepresented among those that tweet scientific journal articles. Tsou, Bowman et al. (2015) found that about a third of Twitter users that had tweeted research articles had a doctoral degree. In another study that surveyed 1,912 people who had tweeted journal articles, it was found that over half (55%) of the respondents belonged to academia (Mohammadi, Thelwall *et al.*, 2018).

This research aims to conduct the first meta-analysis of altmetric data and to combine the results from previous studies and provide a more generalizable understanding of who tweets scientific articles. Understanding who tweets scientific articles is crucial because it allows researchers and institutions to better understand the patterns for dissemination of scholarly information and assess the impact of research beyond traditional academic circles. Central to meta-analysis is quantitative integration of studies with room for addition of goals of critiquing a field of study and identifying key direction for future conceptual, methodological and empirical work (Card, 2015, p. 17). A meta-analysis allows for the systematic synthesis and comparison of findings from multiple earlier studies. Integration involves either combining effect sizes of studies or comparing them if the effect sizes found in studies systematically differ depending on study characteristics (*ibid*). By pooling data from earlier studies, meta-analysis enables us to identify patterns, trends, and discrepancies across different studies, while providing a more comprehensive understanding of the factors influencing who tweets scientific articles. Our chosen approach also enhances statistical power, enabling more robust conclusions to be drawn regarding the characteristics of individuals who engage in tweeting academic research. In addition, our approach will involve a novel and comprehensive user recategorization that can better accommodate the categorizations used in earlier studies and that can be used in future studies.

# 4 Method

To answer the question of "who" are tweeting academic publications, studies that offered statistics about categorization of Twitter users posting academic publications were included in our research. Both academic journal articles and relevant conference and workshop papers were considered eligible for the current work.

## *4.1 Review framework*

To ensure transparency, reproducibility, and comparability, this study followed systematic review and meta-analysis guidelines. The PRISMA checklist (Page, McKenzie et al., 2021) was used to structure the review, and the GRADE framework (Schünemann, Brożek et al., 2013) was adopted for evaluating the quality of evidence. The protocol was designed to facilitate future continuous reviews, especially given the diversity of various social media users and evolving nature of engagements across platform like X/Twitter. Figure 1 gives an overview of the steps taken in the study.

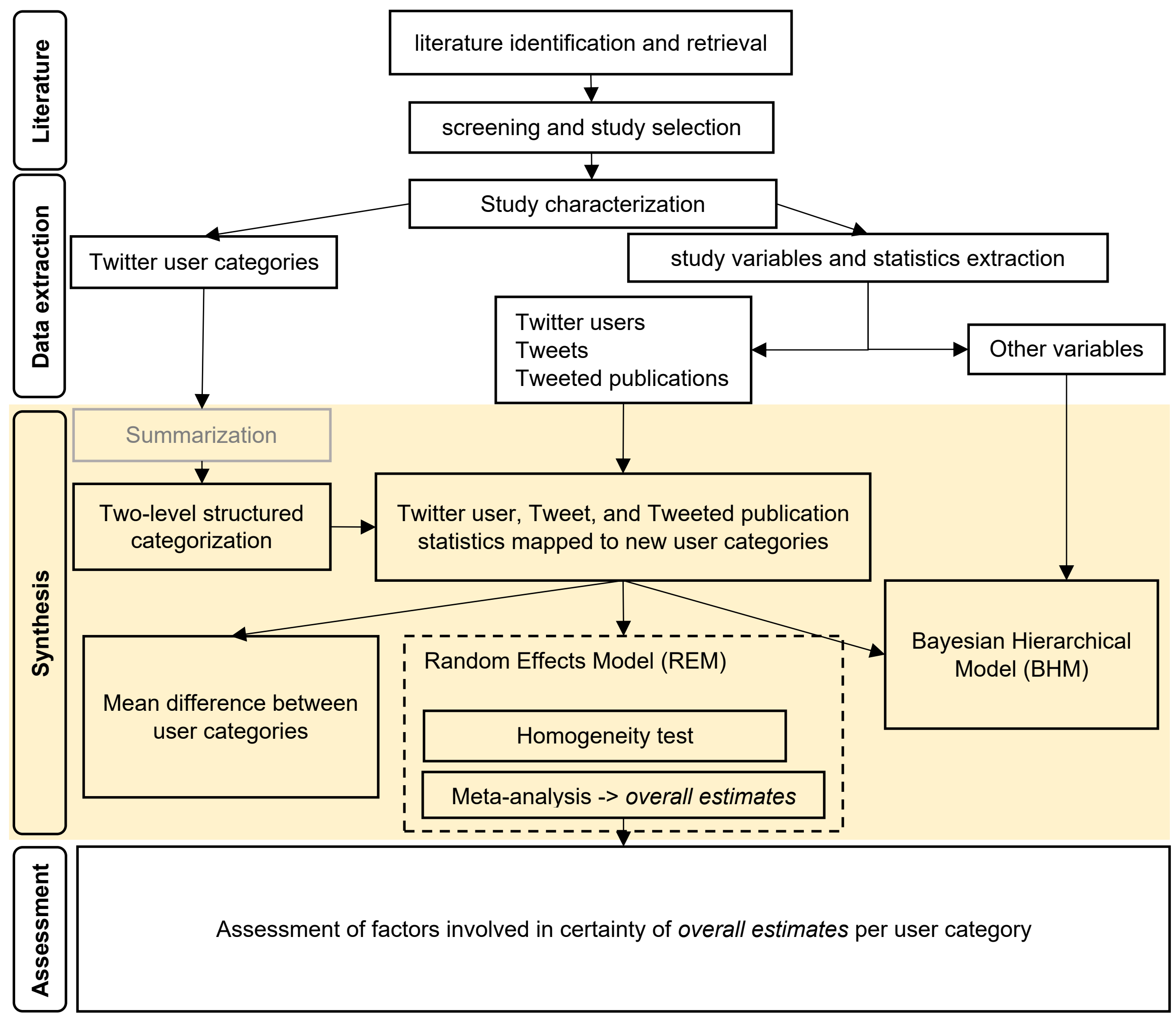

***Figure 1. The steps taken for systematic review and meta-analysis of literature on contribution of different categories of Twitter users in dissemination of scientific publications.***

## *4.2 Data collection and recategorization*

The search terms "altmetric" and "Twitter" or "tweet" were entered into five academic publication databases (Web of Science, Scopus, Google Scholar, PubMed, Dimensions, and OpenAlex), two publisher websites (Wiley and Springer Nature), and the Altmetric workshops during November 2021 (more in Figure 3). In addition to browser searches, the software Publish or Perish (Harzing, 2007) was also used, and the search results were merged. The first round of screening was based on the titles and abstracts of the results. The second round of screening consisted of checking the full text of publications that reported some statistics at the Twitter user level, and data extraction was performed on those that met the criteria. The articles that contained a *Twitter user categorization* and enough information to conduct a meta-analysis on the statistical tests were then used for further analysis. Meta-analysis can be done with at least two studies (Valentine, Pigott & Rothstein, 2010), but technically the power of tests in meta-analysis depends on a number of factors (Pigott, 2012, p. 37). Earlier studies had taken different approaches for user categorization, that also influenced the research design of the present study. As the earlier studies had collected their data with Twitter users, tweets, or tweeted publications, as a starting point, these three approaches were also used here to group the data from the chosen earlier studies. Thus, all the analyses were conducted separately on the studies in these groups:

Twitter users, tweets, or tweeted publications. Relevant statistics were extracted from the content, or the tables and figures presented in the articles. Due to varied user categorizations across studies, we reclassified user types (Table 1) into a unified two-level structure:

(1) Unit level – Individual, Organization, Science Communicator; and

(2) Function level – Academic, Professional, Bot, Mixed/Other.

An additional Mixed Groups category was used for undefined or ambiguous users. This process reduced 376 categories (Supplemental table S1) across studies to 11 harmonized categories.

The category of *Individuals excluding Academics* was later added to make comparisons with *Academic Individuals*. This categorization prioritized the most significant categories while also accommodating both broad and detailed categorization approaches across the two levels, thereby enabling meta-analysis to encompass a wide range of studies.

***Table 1. Twitter user categories as in our recategorization with examples from the earlier studies (the new categories in bold, the abbreviations in parentheses, and examples of categories used in earlier studies in normal text).***

| *User Units (ownership)* | *User Functions (professional associations)* | | *Examples / definition* |
|---|---|---|---|
| **Individuals (I)** | **Academic (A)** | | scientist, researcher, faculty member |
| | **Non-Academic or Individuals excluding Academics ($I\not\subset A$)** | **Professional (P)** | Practitioner, Entrepreneur |
| | | **Mixed / Other (M)** | student, citizen, non-academic, unaffiliated, unspecified individual |
| **Organizations (O)** | **Academic (A)** | | university, research institutions |
| | **Professional (P)** | | enterprise, corporation, association, business |
| | **Mixed / Other (M)** | | organizational accounts |
| **Science communicators (SC)**: accounts systematically disseminating scientific articles | **Academic (A)** | | library, librarian |
| | **Professional (P)** | | journal, publisher, media/news/journalist |
| | **Bot (B)** | | bot, fully automated bots |
| | **Mixed / Other (M)** | | partially-automated bots, science communicator |
| **Mixed groups (MG)** | **Mixed groups (MG)** | | unspecified accounts and a mix of unit level accounts: member of the public, non-bot, non-academic, non-professional, unknown, undefined, unable to tell |

Categorization variability across studies led to possible overlaps and inconsistencies. For example, practitioner researchers could appear in both academic and professional groups to due their hybrid roles as practicing researchers. Similarly, science communicators could include individuals, bots, and media outlets, making clear boundaries difficult. Some user roles (e.g., cyborgs in Haustein, 2019) were excluded due to limited representation. Overlaps and inconsistencies were mitigated through structured reclassification and separate treatment of ambiguous or hybrid user types (such as in Haustein, 2019 and Didegah, Mejlgaard, & Sørensen, 2018). See the established structure at study level in Table 5 in Findings.

Only three key metrics were analyzed across studies: (1) count/proportion of Twitter users, (2) count/proportion of tweets, and (3) count/proportion of tweeted publications. These were derived from user-linked count data, either directly from studies or calculated from reported values. Other metrics, such as follower counts or median tweets, were excluded due to insufficient data across studies.

Additional variables were extracted to assess presence of systematics sources of heterogeneity and are used as binary variables in Beta-Binomial Hierarchical Model to account for the variability across studies. These variables included details such as the method of coding Twitter users (manual or automated), number of coders in the manually coded studies, sample size of each subject of measurement (count of publications, tweets, and Twitter users), information about the tweeted publications (subject field, source of publication including database and journal, publication year, and geographic location of author), and tweets (collection date, types of tweets, and their content). Table 2 provides a list of these factors and their subgroups with definitions.

***Table 2. List and definition of other variables examined in BBHM meta-analysis***

| Factors | Subgroups | Definition |
|---|---|---|
| ***Twitter User Coding Method*** | Manual | studies with manual Twitter user categorization |
| | Automated/Manual-Automated | studies with automated Twitter user categorization |
| ***No of user categories*** | <=4 categories | Studies with broad and not more than 4 user categories |
| | 4< categories | Studies offering over 4 or detailed Twitter user categorization |
| ***Sample Size*** | <1,000 | sample sizes of tweeted publications or Twitter users below 1,000 |
| | 1,000< | sample sizes of tweeted publications or Twitter users above 1,000 |
| | <10,000 | sample sizes of tweets below 10,000 |
| | 10,000< | sample sizes of tweets above 10,000 |
| ***Activity level*** | High activity | Studies on highly tweeting users, or highly tweeted publications or fields |
| | Mixed | Studies covering a mix of dataset or not specifying activity level |
| ***Number of Fields*** | Single subject | studies on publications in one subject field |
| | Multiple subject fields | studies on publications in more than one subject field |
| | Not specified | studies on publications without subject field specification |
| | Multiple subject fields + unspecified | studies on publications in more than one subject field or without subject field specification |
| ***Locations*** | One country/institution | studies on publications of authors from one country or institution |
| | Not specified | studies on publications without author location specification |
| ***Number of Publication Years*** | One year | studies on tweeted articles published in a single year |
| | A range of years | studies on tweeted articles published in a range of years |
| ***Twitter Data Collection Year*** | 2013-2015 | The year Twitter mentions to publications have been collected |
| | 2016-2018 | |
| | 2019-2021 | |

## *4.3 Synthesis method*

The synthesis aimed to identify cross-study patterns in the involvement of different Twitter user types in disseminating scholarly content. The proportions of Twitter users, tweets, and tweeted publications served as the primary outcome variables. Extracted metrics were standardized by recalculating proportions based on reported user counts relative to sample sizes. When studies presented results across multiple topics (Zhou & Na, 2019) or time periods (Xia, Su et al., 2016), a single representative metric per study was retained to avoid bias. In cases where fine-grained user categories needed merging, values were combined to align with our recategorized classification scheme. Figure 2 illustrates the process used to calculate the percentages of tweets and Twitter users.

When studies used differing datasets or coding schemes (e.g., Abhari, Vincent et al., 2022, now as Dambanemuya, Abhari et al., 2022; Ye, Na, & Oh, 2022), they were treated as separate datasets. Some studies reported proportions of tweeted articles (such as Didegah, Mejlgaard, & Sørensen, 2018; Maleki, 2016), while others reported on the proportion of all articles, including articles with zero tweets (Xia, Su et al., 2016; Maleki, 2018). For tweeted publications, only records with at least one tweet were analyzed. Zero values were not imputed for missing categories.

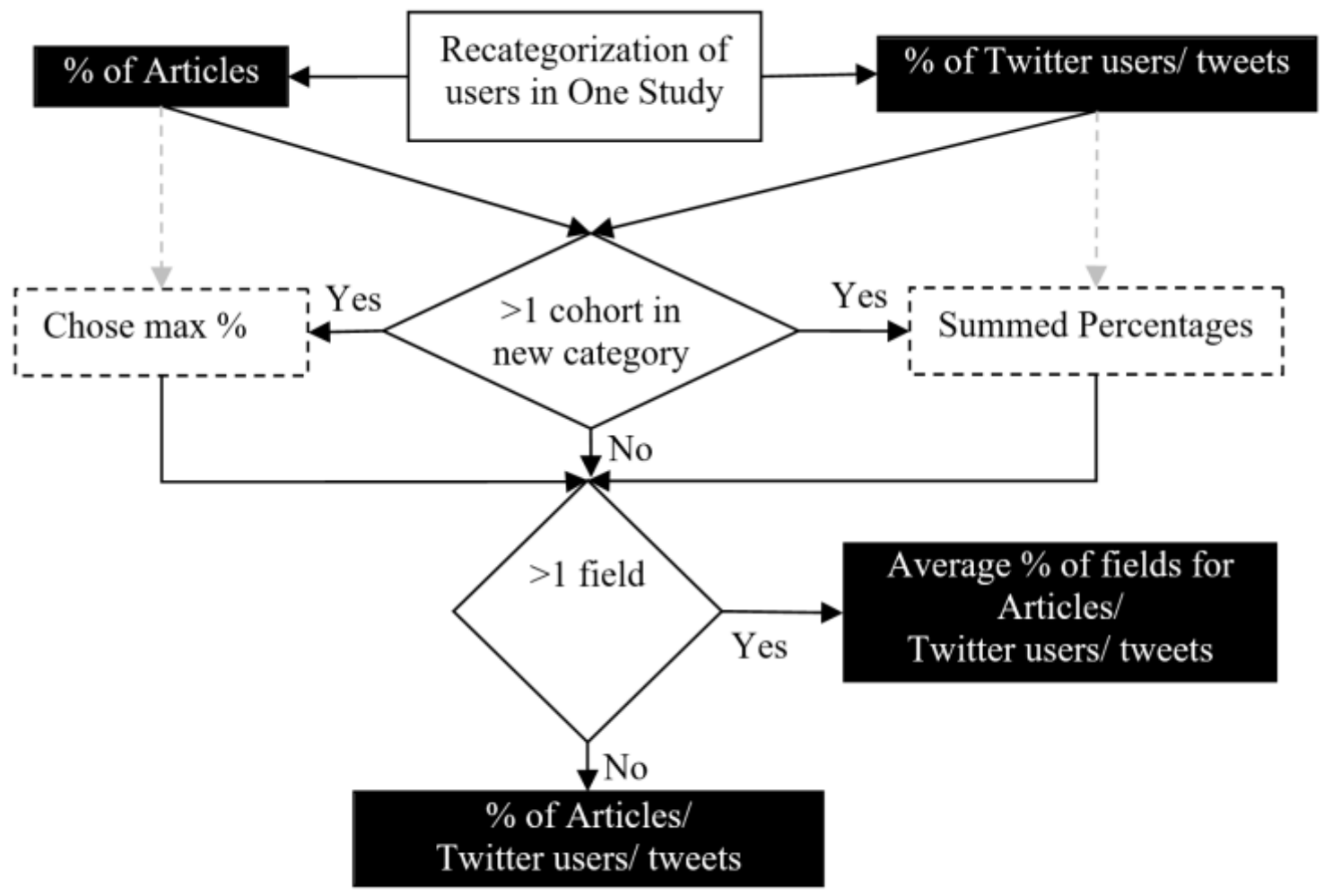

***Figure 2. The process scheme for choosing and calculating the statistics based on the new user categorization***

**Homogeneity Test**. To assess statistical consistency, Cochran's Q test and $I^2$ statistics were used. A p-value $< 0.05$ and $I^2 > 50\%$ were interpreted as indicators of substantial heterogeneity, justifying the use of random-effects models (Mikolajewicz & Komarova, 2019). $H^2$ was calculated using $H^2 = Q_{total}/df$, and $I^2$ using $I^2 = (Q - df) / Q$.

## 4.4 Meta-analysis Approach

### 4.4.1 Random-Effects Model (REM)

A random-effects model using Restricted Maximum Likelihood (ReML) estimation was used to calculate pooled estimates (Bartlett, 1937). Random effect models is assumed to represent a random sample from a particular distribution of the effect sizes and assist in controlling for unobserved heterogeneity (Wooldridge, 2010). REM has historical relevance in meta-analysis studies and could be useful if future research needs to reference its estimates.

*Study weights*, $w_i$, in REM are calculated using the inverse variances derived from *standard error* plus *between-study variance* or *heterogeneity,* $\tau^2$ (Mikolajewicz & Komarova, 2019). Because most of the earlier studies did not provide the *standard deviation*, the *standard error*, $se(\theta_i)$, is used, which is approximately the inverse root of the study-level *sample size,* $n_i$. Study weights ($w_i$) were derived using:

$$se(\theta_i) \sim \frac{1}{\sqrt{n_i}}$$

$$w_i = \frac{1}{se(\theta_i)^2 + \tau^2}$$

In the current study, except for the *standard error*, other measures such as *study weights*, and *confidence intervals* were derived and illustrated using SPSS version 28.0.0.0 (IBM, 2021). All the meta-analysis estimates were made in 95% confidence interval. Forest plots were used for visualizing study estimates and their confidence intervals.

Meta-regression was initially considered to evaluate the impact of study-level variables. However, due to limited data and sparse subgroup representation, it yielded unstable and non-generalizable estimates. Consequently, meta-regression was not pursued. Instead, binary-coded study-level variables were embedded as weakly informative priors in a Bayesian hierarchical model. This allowed us to incorporate potential sources of heterogeneity without inflating error margins.

Traditional REM and *proportion-weighted* approaches are constrained by their sensitivity to large-sample outliers and sparse category coverage. *Weighted means*, for example, disproportionately favor studies like Yu (2017), which reported over 5 million tweeted publications, and can obscure meaningful variation from smaller studies. BBHM addresses these limitations by modeling engagement as a probabilistic distribution rather than a fixed point estimate. It accommodates between-study variance through hierarchical priors and fully propagates uncertainty, particularly critical when category representation is low. It also treats the engagement outcomes as count data, making it compatible with the binomial structure of the problem. This model is ideal for scenarios like ours, where user classification is non-exhaustive, overlapping or not mutually exclusive, and sometimes ambiguous. BBHM simultaneously estimates the probability of a user being assigned to a category and not assigned to that category, enabling clearer insights into classification dynamics.

### 4.4.2 Beta-Binomial Hierarchical Model (BBHM)

To address high heterogeneity, variable sample sizes, and sparse studies, we employed a Beta-Binomial Hierarchical Model (BBHM).

Let $x_i$ be the number of classified users in a category and $n_i$ be the total sample size in study $i$. The Beta-Binomial Hierarchical Model (BBHM) approach allows engagement outcomes to be modeled as binomial events:

$$x_i \sim \text{Binomial}(n_i, p_i)$$

where $p_i$ represents the study-specific user engagement probability and is modeled as a Beta-distributed random variable:

$$p_i \sim \text{Beta}(\alpha_i, \beta_i)$$

Here, α reflects support for user identification in a given user category ("in-category"), while $\beta_i$ represents exclusion ("out-of-category"). In this way, BBHM provides a natural framework for interpreting both the likelihood of a user being classified into a specific category and the complementary probability of not being classified into that category. This modeling strategy captures uncertainty and variability in user classification performance across studies.

At the user category level, the BBHM yields a posterior estimate $\rho$, defined as the mean of the beta distribution from which each study-specific probability $p_i$ is drawn. This value represents the overall expected engagement proportion for a given user category across all studies, accounting for between-study variability. In other words, $\rho = \alpha / (\alpha+\beta)$ reflects the model's best estimate of how frequently users in that category engage with scholarly publications on Twitter. This estimate, along with its *highest density interval (HDI)*, offers a probabilistic summary of category-level engagement based on all available evidence.

BBHM was implemented in Python using PyMC (draws = 100000, chains=4 and target_accept=0.95) with other binary variables given as weakly informative priors and four Markov Chain Monte Carlo (MCMC) chains of 100,000 draws. Monte Carlo standard error (MCSE) was used to assess sampling precision. Posterior estimates were interpreted as probabilistic engagement rates, allowing category-wise comparisons under a unified modeling structure.

## *4.5 Mean Difference Between User Categories*

To assess differences in engagement across user categories, one-tailed and two-tailed t-tests were performed on study-level proportions. This allowed for comparative evaluation of academic individuals versus other groups across all three engagement metrics.

### 4.6 Certainity of Evidence Assessment

Following GRADE guidelines, results were rated for certainty based on risk of bias, imprecision, inconsistency, indirectness, and publication bias (Schünemann, Brożek et al., 2013). Automated coding studies were considered less reliable than manual ones, and mixed user categories were rated lower due to definitional ambiguity. Imprecision was judged based on confidence interval width and the number of contributing studies. Inconsistency was measured via heterogeneity indicators, and indirectness by relevance of categories to the core research question. Publication bias was assessed qualitatively.

# 5 Findings

## 5.1 Study selection

A comprehensive search was conducted across major academic databases, publisher websites, and the Altmetric.org platform, yielding 3,760 initial records. After removing 1,138 duplicates, 2,622 studies remained for screening. Of these, 1,429 were unrelated to altmetrics, 910 were not relevant to the research question, and 140 were non-English. After applying inclusion criteria, 65 studies were selected for data extraction.

From these, 23 studies contained sufficient quantitative information on Twitter user categorization to be included in the meta-analysis (Figure 3). Several studies were excluded for methodological or conceptual reasons. These were studies that focused on only a single user group (e.g., scientists as in Holmberg & Thelwall, 2014; Ke, Ahn, and Sugimoto, 2017 or journal accounts as in Kwak & Lee, 2014; Said, Bowman et al., 2019), or those that used incompatible categorization schemes – such as merging individuals and organizations (e.g., Pandian, Devi et al., 2019), or applying unusual identification methods like scientists' name-matching (e.g., Joubert and Costas, 2019). Some datasets exhibited high statistical heterogeneity or presented non-comparable results and were only retained after normalization or excluded entirely.

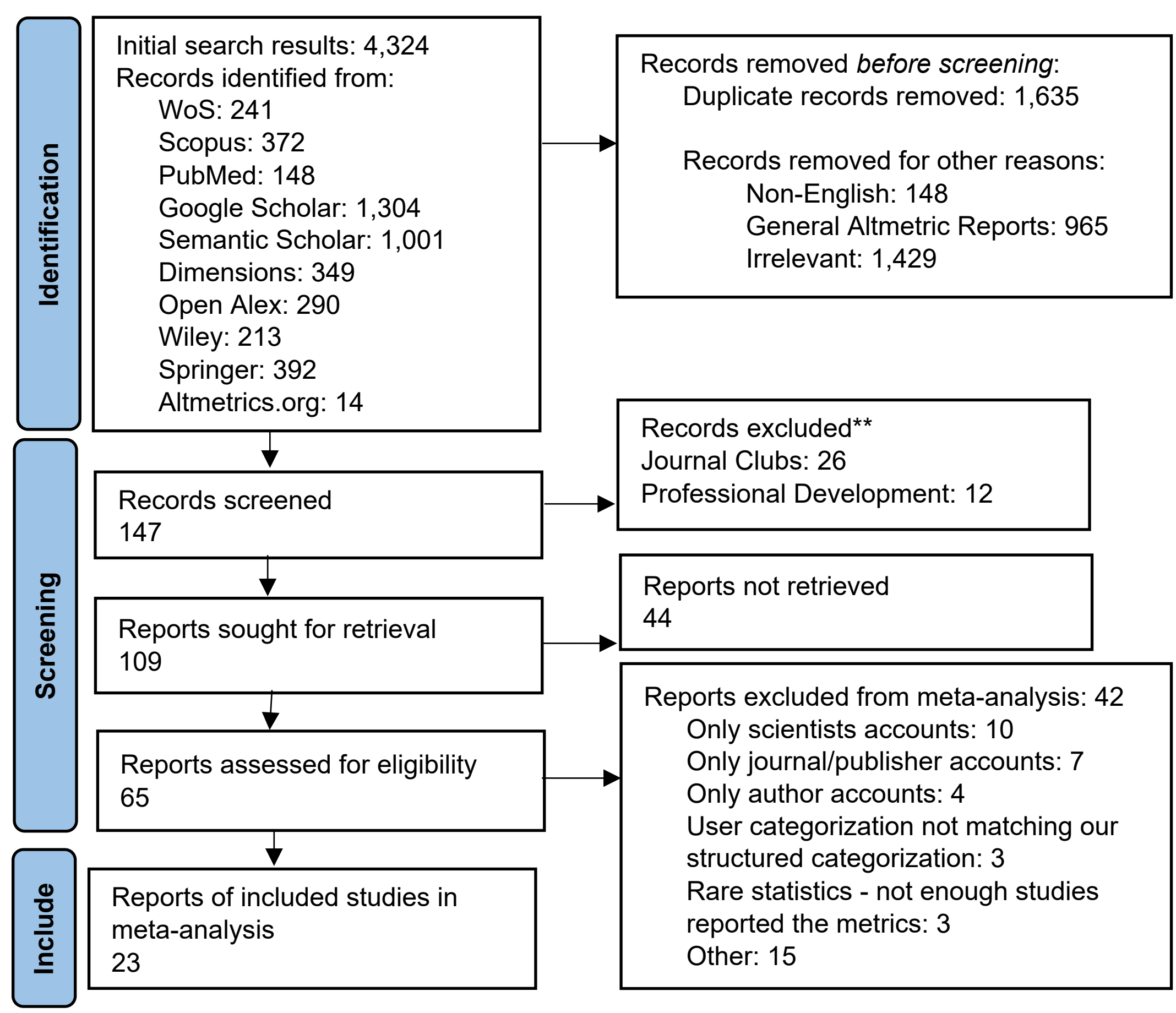

***Figure 3. Flow diagram showing the results of the literature search and the screening of Twitter users studies in Altmetrics and their selection process for systematic review and meta-analysis.***

## 5.2 Study descriptions

The final 23 studies reflected a diverse set of research designs, subject areas, and temporal coverage. Nine studies focused on specific topics such as climate change (Toupin, Millerand & Larivière, 2019) or COVID-19 (Ye, Na, & Oh, 2022), while psychology emerged as the most frequently studied field (Htoo & Jin-Cheon, 2017; Udayakumar, Senadeera et al., 2018; Zhou & Na, 2019). Some studies were geographically bounded, focusing on publications from Finland (Vainio & Holmberg, 2017), Iran (Maleki, 2014), or Brazil (de Melo Maricato & de Castro Manso, 2022).

A few studies analyzed institutional or journal-based tweeting patterns (e.g., Tsou, Bowman et al., 2015), while others examined specific datasets such as arXiv (Haustein, Bowman, 2016) or university repositories (Sergiadis, 2018). Two large-scale studies (Haustein, 2019; Yu, Xiao et al., 2019) analyzed millions of tweets without topical or regional restrictions.

Seven studies targeted high-activity zones on Twitter, including trending disciplines (e.g., astronomy and COVID-19), journals (e.g., *Nature*), or prolific users. While these studies offered rich datasets, their focus on highly active profiles (Haustein et al., 2016; Haustein, 2019; Yu, Xiao et al., 2019) may have introduced sampling biases, resulting in underrepresentation of lesser-active user types. Table 3 provides an overview of study-level characteristics, while excluded studies are detailed in Table 4 in accordance with PRISMA guidelines.

### *Table 3. The data characteristics of the earlier studies included in the meta-analysis.*

| ID | Study | (# of) Coder | Twitter user categorization | User category-based metrics | Dataset | | | Publication Dates | Data collection date | Subject field | Geography |
|---|---|---|---|---|---|---|---|---|---|---|---|
| | | | | | Twitter users | Tweets | Publications | | | | |
| 1 | Maleki 2014 | One | Manual (10 cohorts) | Tweeted articles<br>Tweets | 565 | 1,096 | 589 | 1997-2012 | July 2013 | 5 broad fields | Iran |
| 2 | Tsou 2015 | Two | Manual (12 cohorts) | Twitter users | 2,000 (500 per journal) | - | - | - | March 2012-March 2013 | 9 fields | - |
| 3 | Haustein Bowman 2016 | Two | Manual (3 cohorts) | Tweets | 51 accounts (mentioned arXiv in profile & tweeted) | 50,068 | - | 2012 | May 2014 | - | - |
| 4 | Haustein Tsou 2016 | NR | Manual (22 cohorts) | Twitter users | 800 (200 per activity level) | 663,547 | - | 2012 | June 2015 | - | - |
| 5 | Maleki 2016 | Altmetric.com | 4 cohorts | Tweeted articles | - | - | 8,978 articles | 2011-2014 | August 2016 | - | - |
| 6 | Xia 2016 | Altmetric.com | 4 cohorts | Articles | - | - | 4,276 – has duplicates | 2010-2015 | June 2015 | 4 fields in Nature | - |
| 7 | Htoo 2017 | Two | Manual (4 cohorts and 27 sub-types) | Tweets<br>Twitter users | 1,452 | 2,016 | 133 highly tweeted articles | 2013 | NR | Psychology | - |
| 8 | Vainio 2017 | Two | Manual (16+1[a] cohorts) | Twitter users | 400 (100 per field) | - | 20 articles (top 5/field) | 2014-2016 | 2015 | 4 broad fields | Finland |
| 9 | Yu 2017 | Altmetric.com | 4 cohorts | Tweeted articles<br>Tweets | - | 21,255,678 | 5,184,525 | 2011-2016 | 2016 | 27 fields | - |
| 10 | Didegah 2018 | Three | Manual (14 cohorts) | Tweeted articles<br>Tweets | - | 6,388 | 300 (60 per field) | 2014 | June 2016 | 5 fields | - |
| 11 | Maleki 2018 | Altmetric.com | 4 cohorts | Tweeted Articles | 31,051 | 150,893 | 29,543 articles | 2011-2017 | 2018 | Astrophysics | - |
| 12 | Sergiadis 2018 | One | Manual (6 cohorts) | Tweets | - | 605 | 290 | 1978-2016<br>Mostly in 2012-2016 | October 2016 | 8 fields | Georgia Southern University |
| 13 | Udayakumar 2018 | Two | Manual and Classifier (6 cohorts) | Twitter users | 3,000 | 95,088 | 37,627 | 2014-2015 | NR | Psychology | - |
| 14 | Haustein 2019 | One | Manual (2+1[a] cohorts) | Tweets<br>Twitter users | 2,043 users tweeting ≥1000 | 24.3 million | 3.9 million | 2015 | June 2016 | - | - |
| 15 | Toupin 2019 | Two | Manual (7 cohorts)<br>Semi-automatic | Twitter users | 17,837 of 21,965 | 41,019 | 2,620 of 4,719 | 2015-2016 | NR | Climate change | - |
| 16 | Yu 2019 | Three | Manual (10 + 17 cohorts) | Twitter users | 1,468 (about 500 per activity level in 2016) | 88,407 of sampled tweeters | - | - | October 2011-June 2016 | - | - |
| 17 | Zhou 2019 | Two | Manual (3 x 2 cohorts) and Supervised Learning (15 cohorts) | Twitter users | 6,000 (3,000 per field)<br>33,019 in the test set (10,478 in Politics and 22,541 in Psychology) | 39,532 in Politics and 252,310 in Psychology | 8,265 in Politics and 42,891 in Psychology | 2011-2015 | March 2017 | Psychology & Political Science | - |
| 18 | Haustein 2020 | Two | Manual (11+1[a] cohorts) | Twitter users | 464 | 577 | 25 of 131 | - | October 2018, August 2019 | Judit Bar-Ilan's Publications | - |
| 19 | Lemke 2021 | One | Manual (12 cohorts) | Twitter users<br>Tweets<br>Twitter followers | 50 of 1,808 | 1,161 of 3,677 | - | - | July-October 2020 | German nuclear repository | Germany |
| 20 | Orduña-Malea 2021 | NR | Manual (4 cohorts) | Twitter users | 1,376 (200 originally tweeting, 748 liking, 728 retweeting) | - | - | - | 2019 | VOSviewer software’s website | - |
| 21 | Abhari 2022 | Two | Manual and Rule-based classifier (5 cohorts) | Twitter users<br>Tweets | 26,260 (20,195 retracted + 6,065 non-retracted) | 93,127 (67,124 retracted + 25,997 non-retracted) | 3,847 retracted<br>2,085 non-retracted | 2011-2019 | September 2021 | Retracted articles | - |
| 22 | de Melo Maricato 2022 | NR | Manual (4 cohorts and 36+1 user occupation) | Twitter users<br>Tweeted articles<br>Tweets | 2435 | 3653 | 877 | - | 2020 | 22 fields | University of Brasília |
| 23 | Ye 2022 | Two | Manual (7 cohorts) | Twitter users<br>Tweeted articles<br>Tweets | 659 of 99,619 | 13,179 of 151,480 | 417 of 1,252 | May 2020 | June 2020 | COVID-19 | - |

NR: Not Reported; [a] +1 means one category for all the remaining unclassified users; – in Geography: No geographical restriction

***Table 4. Data characteristics of the earlier studies that were excluded from the meta-analysis***

| ID | Study | Exclusion reason | (# of) Coder | Twitter user categorization | User category-based metrics | Dataset | | | Publication Dates | Data collection date | Subject field |
|---|---|---|---|---|---|---|---|---|---|---|---|
| | | | | | | Twitter users | Tweets | Publications/Authors | | | |
| 1 | Holmberg Thelwall 2014 | One-cohort study | No Twitter user coding | One (researchers) | Tweets | 447 (ranging between 24-52 across fields) | 2000 (200 per field) | - | - | March-October 2012 | 10 fields |
| 2 | Holmberg Bowman 2014[a] | Not only tweeted articles, @user mentiones in tweet content | One | Manual (12 cohorts) | Twitter users | 518 of 11,252 | 27,923 of 56,415 | 32 of 37 Astrophysicists | - | May 2013 | Astrophysics |
| 3 | Kwak 2014 | One-cohort study | No Twitter user coding | One (journals) | Journals holding an account Twitter followers | 57,699 | 136,139 | 31,661 articles in top 100 journals | - | January-June 2013 | 54 domains |
| 4 | Bowman 2015 | One-cohort study | No Twitter user coding | One (Professors surveyed) | Scientists holding an account Personal / Professional usage | 445 users | - | 1910 authors | - | May 2014 | - |
| 5 | Costas 2017 | One-cohort study | No Twitter user coding | One (Authors disambiguated) | Scientists holding a Twitter account | 334,856 | 2,622,116 | 17,332,510 scholars | 1980-2015 | April 2016 | 4 broad fields |
| 6 | Hughes 2017 | One-cohort study | No Twitter user coding | One (journals) | Journals holding a Twitter account | - | - | 50 top journals | - | June 2017 | Trauma and Orthopaedic Surgery |
| 7 | Ke 2017 | One-cohort study | No Twitter user coding | One (Scientists) | Scientists holding a Twitter account Tweets Twitter followers | 45,867 of 110,708 | 64,449,234 | 894,840 scientists in employment | - | May 2014 (Employment data) | 24 fields 5 broad fields |
| 8 | Ortega 2017 | One-cohort study | Targetted groups | 3 cohorts of Publisher, Journal, and Owner | Articles | - | - | 4,176 from 350 journals | 2013 | April 2016 | 5 broad fields |
| 9 | Wong 2018 | One-cohort study | No Twitter user coding | One (journals) | Journals holding a Twitter account Tweets Twitter followers | 18 | - | 50 journals | 2008-2016 | NR | Otolaryngology |
| 10 | Côté 2018 | One-cohort study | No Twitter user coding | One (Scientists) | Twitter followers | 110 of 167 | - | 200 researchers | 2018 | 2015 | Ecology and Evolutionary Biology |
| 11 | Tur-Viñes 2018 | One-cohort study | No Twitter user coding | One (journals) | Journals holding a Twitter account Twitter followers | 30 | - | 70 journals | - | 2016 | Communication |
| 12 | Hassan 2019 | Rare metrics | Altmetric.com | 4 cohorts | Classification score for predicting highly cited articles | - | - | - | 2011-2015 | 2017 | LIS |
| 13 | Joubert 2019 | One-cohort study | One rule-based algorithm | WoS disambiguated list of authors matched on Twitter)->Low recall | - | 14,174 of 2,622,116 | - | - | 2010-2016 | June 2016 | - |
| 14 | Pandian 2019 [a] | Unfit for recategorization | Two | Manual (2 cohorts) | Twitter users | 2,064 | - | 2078 | 2017 | NR | Psychology |
| 15 | Said 2019 | One-cohort study | No Twitter user coding | One (journals) | Twitter Network PageRank and Eigenvector | 19 journals And 149,830 | - | 77,757 | 2015 | Publications (2017) Tweets (2016) | |
| 16 | Yadav 2019 | One-cohort study | No Twitter user coding | One (Journal editorial board) | Editors holding a Twitter account | - | - | 240 editorial members of 6 high IF journals | - | July-August 2018 | Obstetrics and Gynecology |
| 17 | Coret 2020 | One-cohort study | No Twitter user coding | One (surgeons) | Scientists holding an account Tweets/Tweet Likes Twitter followers/friends | 188 | - | 3741 surgeons | - | 2018 | General Thoracic Surgery |
| 18 | Hassan 2020 | Rare metrics | Altmetric.com | 4 cohorts | Tweet sentiments | - | 5,341,800 | - | 2011-2016 | 2016 | - |
| 19 | Clavier 2021 | One-cohort study | No Twitter user coding, but academics | Two (Academic ranks) | Scientists holding an account Tweets/followers | 68 (42 active) | - | 162 authors | 2016-2020 | March 2021 | Anaesthesia |
| 20 | Haunschild 2021[b] | Irregular bot statistics | One R package | Automated (tweetbotornot) | Twitter users | 56,266 | 173,187 | 6,433 | 2011-2019 | NR | Opioid |

| ID | Study | Exclusion reason | (# of) Coder | Twitter user categorization | User category-based metrics | Dataset | | | Publication Dates | Data collection date | Subject field |
|---|---|---|---|---|---|---|---|---|---|---|---|
| | | | | | | Twitter users | Tweets | Publications/Authors | | | |
| **21** | Khandelwal 2021 | One-cohort study | No Twitter user coding | One (Academics) | - | 131 | 56,512 | - | NR | January-February 2020 | Development research |
| **22** | Sanchez 2021 | One-cohort study | No Twitter user coding, but academics | Three (Academic ranks) | Scientists holding an account<br>Age of accounts<br>Tweets/followers/friends | 322 | - | 1,107 scientists | 2019 | March 2007-April 2019 | Urban Planning |
| **23** | Wang 2021 | One-cohort study | No Twitter user coding | One (book publishers) | Publishers holding one or multiple Twitter accounts | - | - | 6,258 of 15,454 unique book titles | 2014-2018 | April and May 2019 | - |
| **24** | Zhang 2021 [c] | Unfit for recategorization | NR | Two x 2 (official and popular) | Tweeted articles<br>Tweets | - | 286,650 | 20,310 in 6 journals | 2009-2019 | July 2019 | journals published by the Royal Society |

NR: Not Reported; WoS: Web of Science; BKCI: Book Citation Index; IF: Impact Factor; LIS: Library and Information Science

[a] The statistics on Academics were a mix of individuals and organizations; [b] The findings introduced significant bias to *Bot Science Communicators*' meta-analysis; [c] The definition of cohorts did not fit with the new categorization.

*Table 5. Outcome measures (percentage values and counts) for Twitter users, Tweets, and Tweeted Publications organized by function-level user categories.*

| Study — Unit | Individual | | | Organization | | | Science Communicator | | | | Mixed Groups | Sample size |
|---|---|---|---|---|---|---|---|---|---|---|---|---|
| Function | **Academic** | **Professional** | **Other/Mixed** | **Academic** | **Professional** | **Other/Mixed** | **Academic** | **Professional** | **Bot** | **Other/Mixed** | | |
| **Twitter users** | | | | | | | | | | | | |
| Tsou 2015 | 26.2% | | 49.9% | 3% | 6.7% | 8.9% | | 4.4% | | | 1.1% | 100% |
| | 523 | | 997 | 60 | 134 | 178 | | 87 | | | 21 | 2,000 |
| Haustein Tsou 2016 | 16.7% | 36.1% | 5.5% | 11.9% | 3.2% | 5.0% | 0.1% | 1.2% | 8.0% | 9.5% | 11.7% | 100% |
| | 134 | 288 | 44 | 95 | 25 | 40 | 1 | 10 | 64 | 76 | 93 | 800 |
| Htoo 2017 | 27.5% | | 55.2% | 7.9% | | 9.4% | | | | | | 100% |
| | 400 | | 801 | 114 | | 137 | | | | | | 1,452 |
| Vainio 2017 | 23.8% | 21.5% | 2.0% | 2.8% | 9.5% | 5.0% | 1.3% | 7.8% | | 0.8% | 25.8% | 100% |
| | 95 | 86 | 8 | 11 | 38 | 20 | 5 | 31 | | 3 | 103 | 400 |
| Udayakumar 2018 | 21.4% | | 46.6% | | | 32% | | | | | | 100% |
| | 641 | | 1,399 | | | 960 | | | | | | 3,000 |
| Haustein 2019 | | | | | | | | 14.9% | 12.1% | | 72.9% | 100% |
| | | | | | | | | 305 | 248 | | 1,490 | 2,043 |
| Toupin 2019 | 31.7% | 2.7% | 1.2%[a] | | | 19.5% | | 12.7% | 1.9% | | 30.4% | 100% |
| | 5,651 | 473 | 216 | | | 3,475 | | 2,265 | 335 | | 5,422 | 17,837 |
| Yu 2019 | 44.9% | 15.8% | 22.2% | 2.3% | 9.1% | 16% | 2.5% | 3.2% | 1.2% | 0.6% | | -[b] |
| | 659 | 232 | 326 | 34 | 134 | 235 | 37 | 47 | 18 | 9 | | 1,468 |
| Zhou 2019 | 30.8% | | 45.7% | 7.2% | | 16.5% | | | | | | 100% |
| | 9,073 | | 14,899 | 2,489 | | 6,558 | | | | | | 33,019 |
| Haustein 2020 | 39.4% | 13.8% | | 5.8% | | 7.1% | 17.0% | 26.3% | 1.5% | | 20.5% | -[b] |
| | 183 | 64 | | 27 | | 33 | 79 | 122 | 7 | | 95 | 464 |
| Lemke 2021 | | | 44% | | | 54% | | | | | 2% | 100% |
| | | | 22 | | | 27 | | | | | 1 | 50 |
| Orduña-Malea 2021 | | | 86.5% | | | 11.2% | | | | | 2.3% | 100% |
| | | | 1,207 | | | 138 | | | | | 31 | 1,376 |
| Abhari 2022 Retracted | 54.3% | 17.6% | | | | | | | 2.8% | 12.3% | 13.0% | 100% |
| | 5,166 | 1,674 | | | | | | | 266 | 1,170 | 1,237 | 9,514 |
| Abhari 2022 Not Retracted | 54.6% | 15.6% | | | | | | | 5.3% | 20.3% | 4.2% | 100% |
| | 1,362 | 389 | | | | | | | 132 | 506 | 105 | 2,495 |
| de Melo Maricato 2022 | 29.6% | 12.4% | 26.0% | | | 25.2% | | 1.4% | | 4.0% | 1.5% | 100% |
| | 719 | 306 | 633 | | | 614 | | 32 | | 97 | 36 | 2,437 |
| Ye 2022 | 22.2%[c] | 27.3% | | | | | | 5.5% | 13.2% | 5.2% | 39.9% | 100% |
| | 146 | 180 | | | | | | 36 | 87[a] | 34 | 263 | 659 |
| **Tweets** | | | | | | | | | | | | |
| Maleki 2014 | 13.0% | 17.0% | 22.0% | 11.0% | 3.0% | | | 11.0% | | 23.0% | | 100% |
| | 110 | 189 | 248 | 137 | 30 | | | 126 | | 256 | | 1,096 |
| Haustein Bowman 2016 | | | | | | | | | | 9% | 91% | 100% |
| | | | | | | | | | | ~4,506 | ~45,562 | 50,068 |
| Htoo 2017 | 29.2% | | 47.6% | 14.2% | | 9.0% | | | | | | 100% |
| | 577 | | 937 | 280 | | 177 | | | | | | 1,971 |
| Yu 2017 | 16.5% | 7.5% | | | | | | | | 4.9% | 71.1% | 100% |
| | 3,503,640 | 1,587,158 | | | | | | | | 1,041,701 | 15,108,440 | 21,255,678 |

| Study (Unit / Function) | Individual | | | Organization | | | Science Communicator | | | | Mixed Groups | Sample size |
|---|---|---|---|---|---|---|---|---|---|---|---|---|
| | Academic | Professional | Other/Mixed | Academic | Professional | Other/Mixed | Academic | Professional | Bot | Other/Mixed | | |
| Didegah 2018 | 21.33% | 10.7% | 34.7% | 1.5% | 20.5% | 0.1% | | 5.2% | 21.1% | 0.04% | 6.0% | 100% |
| | 1,363 | 685 | 2,214 | 93 | 1,308 | 6 | | 333 | 1,336 [c] | 3 [d] | 384 | 6,388 |
| Sergiadis 2018 | 14.2% | | 51.7% | | 11.2% | 22.8% | | | | | | 100% |
| | 86 | | 313 | | 68 | 138 | | | | | | 605 |
| Haustein 2019 | | | | | | | | 3% | 7% | | 90% | 100% |
| | | | | | | | | 0.73million | 1.7million | | 21.9million | 24.3million |
| Lemke 2021 | | | 54.3% | | | 45.0% | | | | | 0.7% | 100% |
| | | | 631 | | | 522 | | | | | 8 | 1161 |
| Abhari 2022 Retracted | 18.2% | 7.7% | 23.8% | | | | | | 8.3% | 5.3% | 36.7% | 100% |
| | 4,933 | 2,079 | 6,435 | | | | | | 2,250 | 1,427 | 9,949 | 27,073 |
| Abhari 2022 Not Retracted | 35.2% | 6.3% | 6.5% | | | | | | 4.9% | 21.1% | 25.9% | 100% |
| | 4,050 | 729 | 750 | | | | | | 564 | 2,723 | 2,984 | 11,500 |
| de Melo Maricato 2022 | 29.6% | 12.3% | 26.0% | | | 25.1% | | 1.3% | | 4.2% | 1.5% | 100% |
| | 723 | 308 | 638 | | | 617 | | 32 | | 102 | 36 | 2,456 |
| Ye 2022 | 19.8% | 26.7% | | | | | | 6.1% | 17.6% | 6.8% | 40.6% | 100% |
| | 2,613 | 3,523 | | | | | | 801 | 2,320 [c] | 896 | 5,346 | 13,179 |
| **Tweeted publications** | | | | | | | | | | | | |
| Maleki 2014 | 11.4% | 19.4% | 3.4% | 16.6% | 4.1% | | | 17.7% | | 31.9% | 20.9% | - [b] |
| | 67 | 114 | 20 | 98 | 24 | | | 104 | | 188 | 123 | 589 |
| Maleki 2016 | 69.6% | 49.1% | | | | | | | | 49.3% | 97.5% | - [b] |
| | 6,251 | 4,407 | | | | | | | | 4,426 | 8,755 | 8,978 |
| Xia 2016 [e] | 83.9% | 26.4% | | | | | | | | 47.2% | 90.6% | - [b] |
| | 3,100 | 975 | | | | | | | | 1,745 | 3,348 | ~3,694 [e] |
| Yu 2017 | 33.1% | 16.2% | | | | | | | | 16.4% | 85.1% | - [b] |
| | 1,292,686 | 631,927 | | | | | | | | 641,232 | 3,323,066 | 5,184,525 |
| Didegah 2018 | 31.8% | 6.8% | 13.4% | 4.0% | 17.2% | | | 4.8% | 32.0% | | 10.7% | - [b] |
| | 74 | 20 | 40 | 12 | 56 | | | 43 | 96 | | 32 | 300 |
| Maleki 2018 | 50.4% | 2.2% | | | | | | | | 25.1% | 78.5% | - [b] |
| | 14,877 | 649 | | | | | | | | 7,423 | 23,196 | 29,543 |
| de Melo Maricato 2022 | 30.7% | 16.2% | 70.8% | | | 46.3% | | 1.9% | | 6.8% | 3.9% | - [b] |
| | 430 | 226 | 621 | | | 406 | | 26 | | 60 | 34 | 877 |
| Ye 2022 [f] | 6.72% | 2.6% | 13.2% | | | | | 9.6% | 5.3% | | 62.6% | - [b] |
| | 28 | 11 | 55 | | | | | 40 | 22 | | 261 | 417 |

**[a] Calculated based on the proportion of "personal" accounts not overlapping with any other category; [b] Overlapping categories (over 100%); [c] This study is imprecise because it might include academic institutions; [d] The study identifies bots and overlaps with other user categories (excluded when summing up user categories to get the sample size); [e] This study reported statistics including papers with zero tweets. The statistics are estimated by the author based on articles with non-zero tweets; [f] Only if articles were originally tweeted by the cohort.**

## 5.3 Study-level Outcome Measures

Table 5 presents the results from the earlier studies that were included for the meta-analysis. The Twitter user types presented in these were examined across three metrics: (a) *Twitter users* (15 studies), (b) *tweets* (11 studies), and (c) *tweeted publications* (8 studies). As some of the included studied presented more than one type of metrics, the total number of studies exceeds the number of publications (23) included. Table 5 shows the counts and percentages of the three metrics for each of the proposed Twitter user categories at function level. However, other outcome measures that were reported in the earlier studies, such as count, percent, mean, median, geometric mean, minimum, maximum, and different percentiles of tweets (12 studies), Twitter followers (9), Twitter users (5), Twitter friends (4), and similar metrics were not analyzed in this study due to too low number of studies to perform a meta-analysis with.

## 5.4 Synthesis by Random Effect Model (REM)

Figures 4-9 indicate results of 42 meta-analyses (3 variables × 11 function-categories (-3 missing due to few studies) + 3 variables × 4 unit-categories).

### 5.4.1 Meta-analysis of Twitter users

Fifteen studies contributed 138 outcome measures as user counts: 48 at the user unit level and 90 at the function level. From a total of 79,014 Twitter accounts, individuals were the dominant contributors to scholarly tweeting, with an average of 66% (95% CI: 59–74%, $I^2$ = 98.9%) (Figure 4). Organizations followed with 22% (CI: 18–25%, $I^2$ = 92.8%), and science communicators 16% (CI: 9–23%, $I^2$ = 97.4%).

At the function level (Figure 5), academic individuals represented 33% of Twitter users (CI: 26–39%, $I^2$ = 98.9%). Academic and professional organizations each had 7% of users (both $I^2$ < 50%, high certainty). Other categories showed lower proportions and high heterogeneity, e.g., professional individuals at 18%, mixed individuals at 28% ($I^2$ = 99.9%), and bot science communicators at 5%.

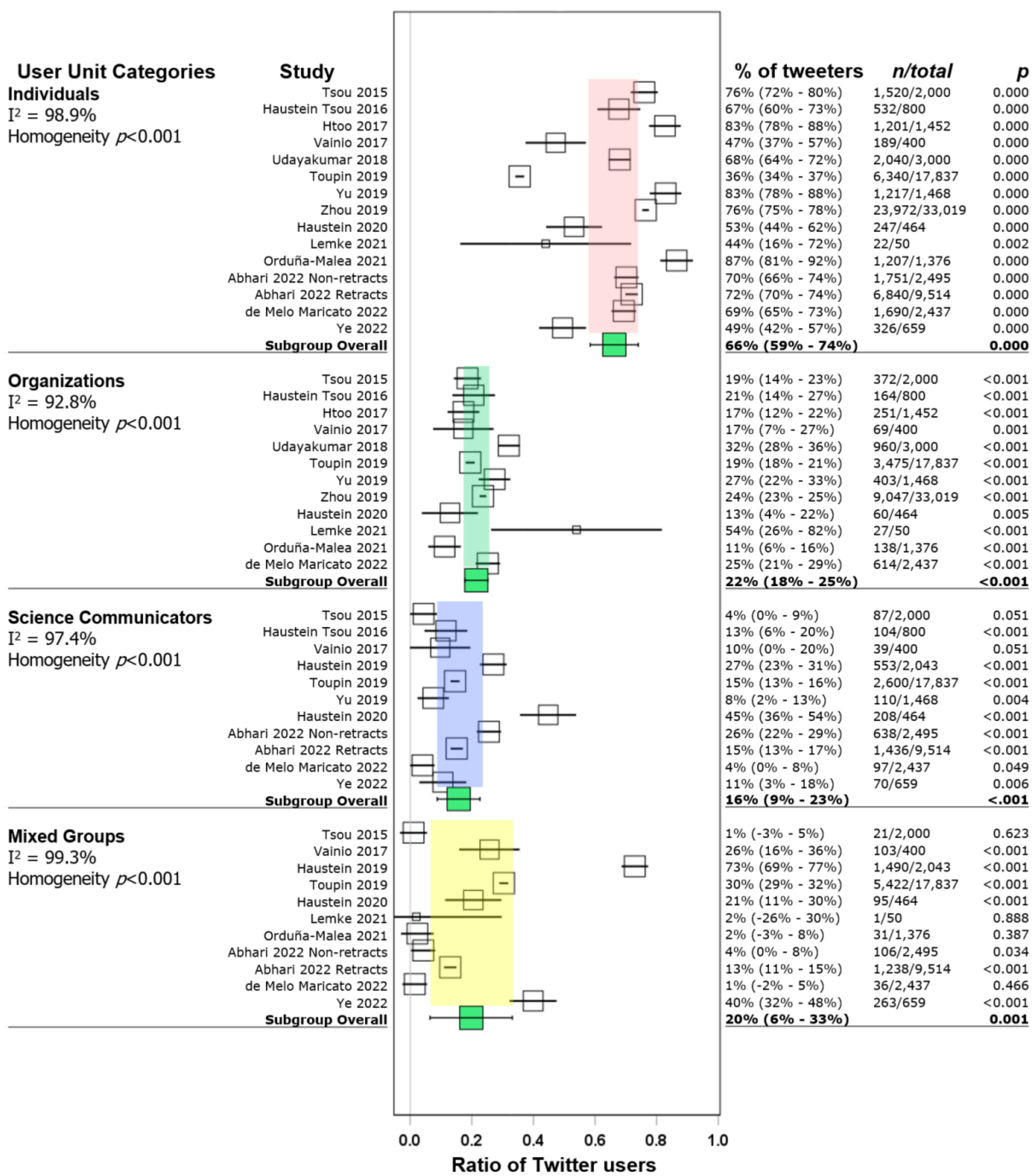

***Figure 4. Meta-analysis of proportion of users at User Unit level. The colour ribbons identify the (non-)overlapping studies with the confidence intervals of the overall estimates for each user category.***

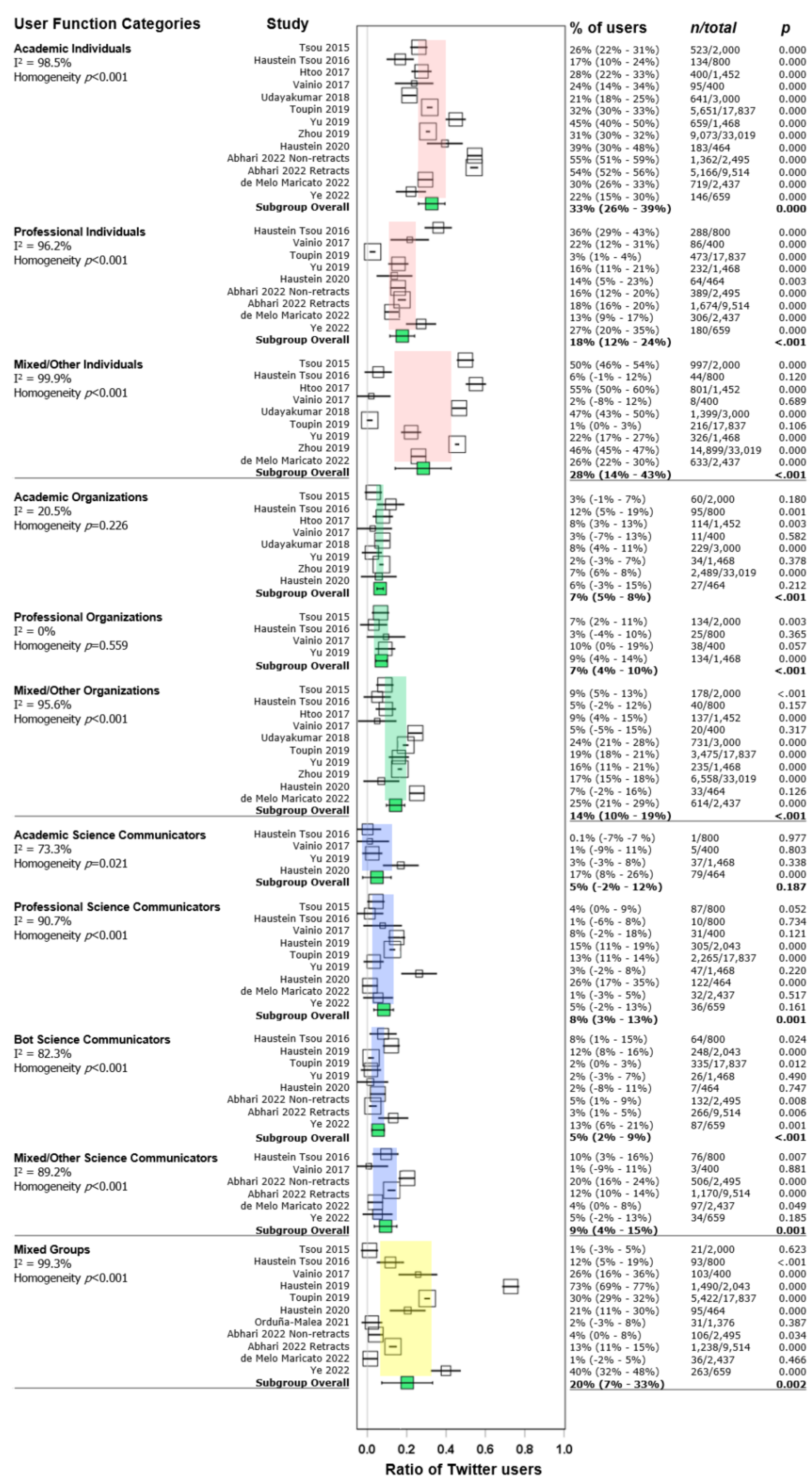

***Figure 5. Meta-analysis of proportion of Twitter users at User Functio level. The colour ribbons identify the (non-)overlapping studies with the confidence intervals of the overall estimates for each user category.***

### 5.4.2 Meta-analysis of Tweets

Eleven studies contributed 92 outcomes, covering a total of 104,112 tweets from nine small-scale studies and over 2 million tweets in two large-scale studies. Individuals produced the largest share of tweets (55%, CI: 46–65%, $I^2 = 99.7\%$), followed by organizations (27%, CI: 19–36%, $I^2 = 95.4\%$) and science communicators (13%, CI: 7–20%, $I^2 = 100\%$) (Figure 6).

At the function level (Figure 7), academic individuals tweeted 22% of the research content (CI: 17–27%, $I^2 = 98.9\%$), and mixed individuals 30% (CI: 19–42%, $I^2 = 99.2\%$). Bot science communicators were below the overall average at 9%, while professional science communicators were the only group with higher certainty (5%, $I^2 = 83.7\%$).

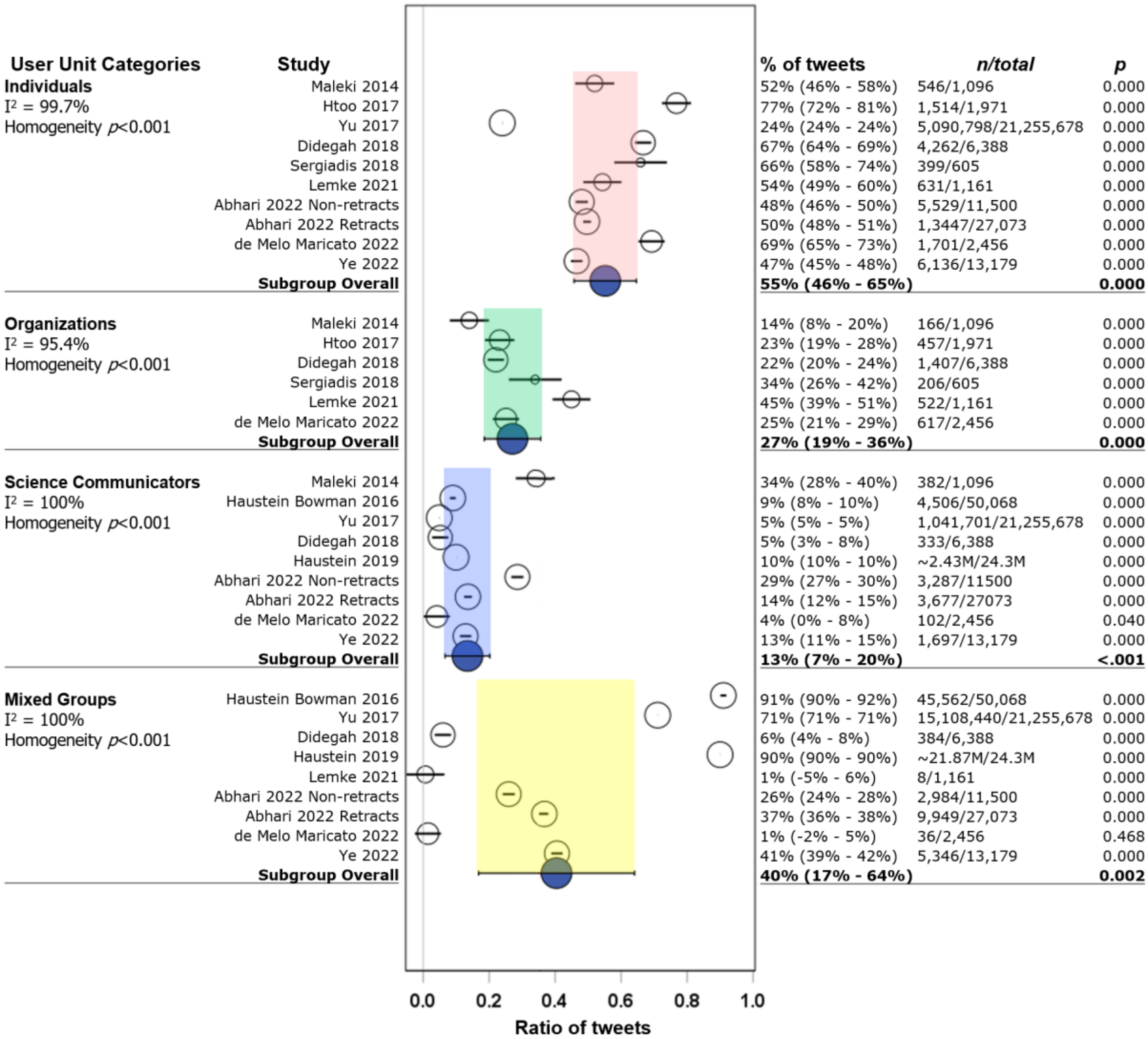

*Figure 6. Meta-analysis of proportion of tweets at User Unit level. The colour ribbons identify the (non-)overlapping studies with the confidence intervals of the overall estimates for each user category.*

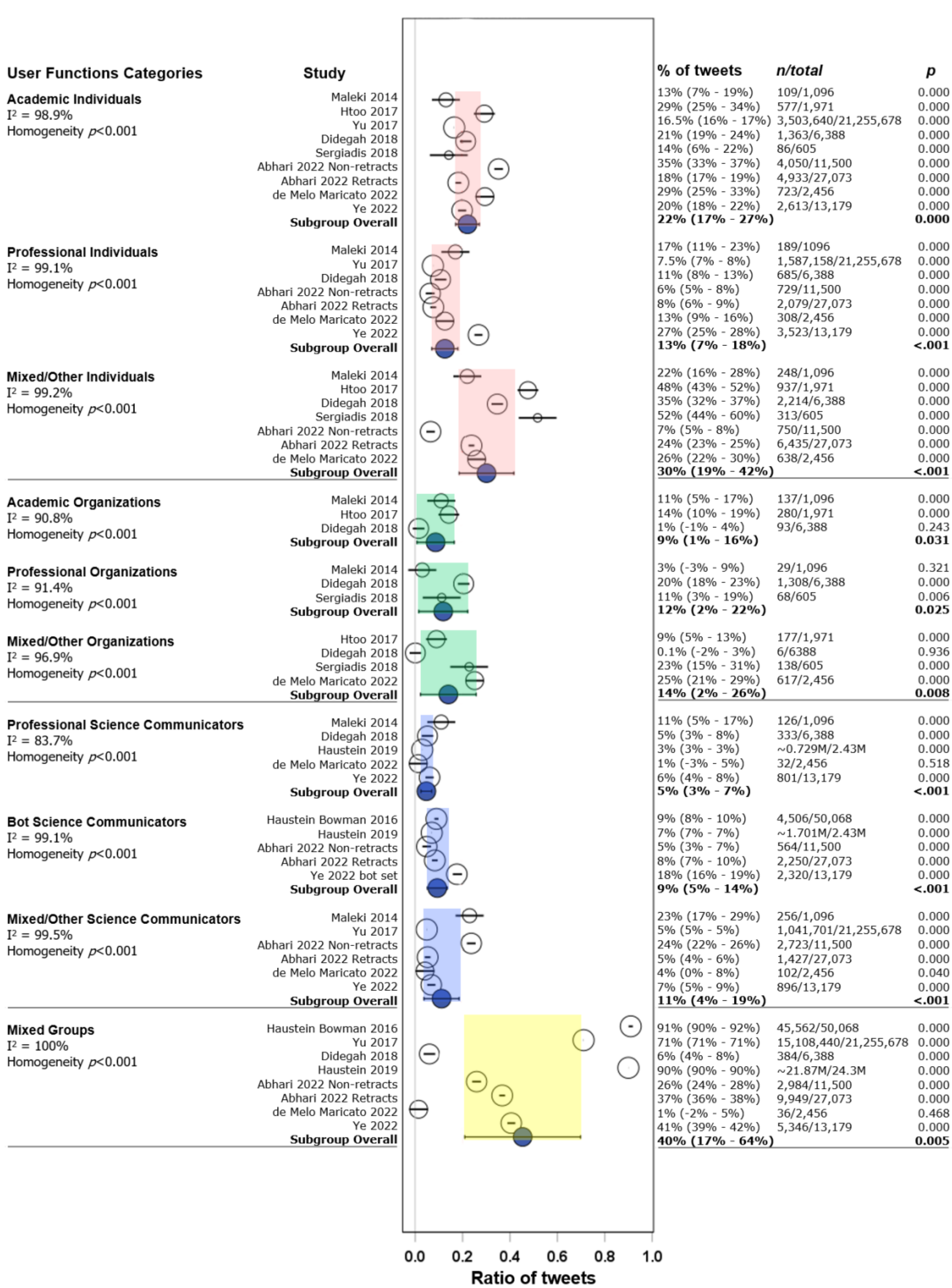

*Figure 7. Meta-analysis of proportion of tweets at User Function level. The colour ribbons identify the (non-)overlapping studies with the confidence intervals of the overall estimates for each user category.*

### 5.4.3 Meta-analysis of Tweeted Publications

Eight studies provided 57 outcomes for tweeted publications: 21 at unit level, and 36 at function level. From 44,398 publications (plus over 5 million from Yu, 2017), individuals were linked to 50% of tweeted articles (CI: 34–66%, $I^2$ = 99.8%), organizations 28% (CI: 10–47%, $I^2$ = 92.8%), and science communicators 30% (CI: 18–42%, $I^2$ = 99.7%) (Figure 8).

At the function level (Figure 9), academic individuals tweeted 42% of publications (CI: 24–61%, $I^2$ = 99.9%). Professional science communicators tweeted 9% (CI: 2–16%, $I^2$ = 61%), with other groups showing high heterogeneity and wider uncertainty.

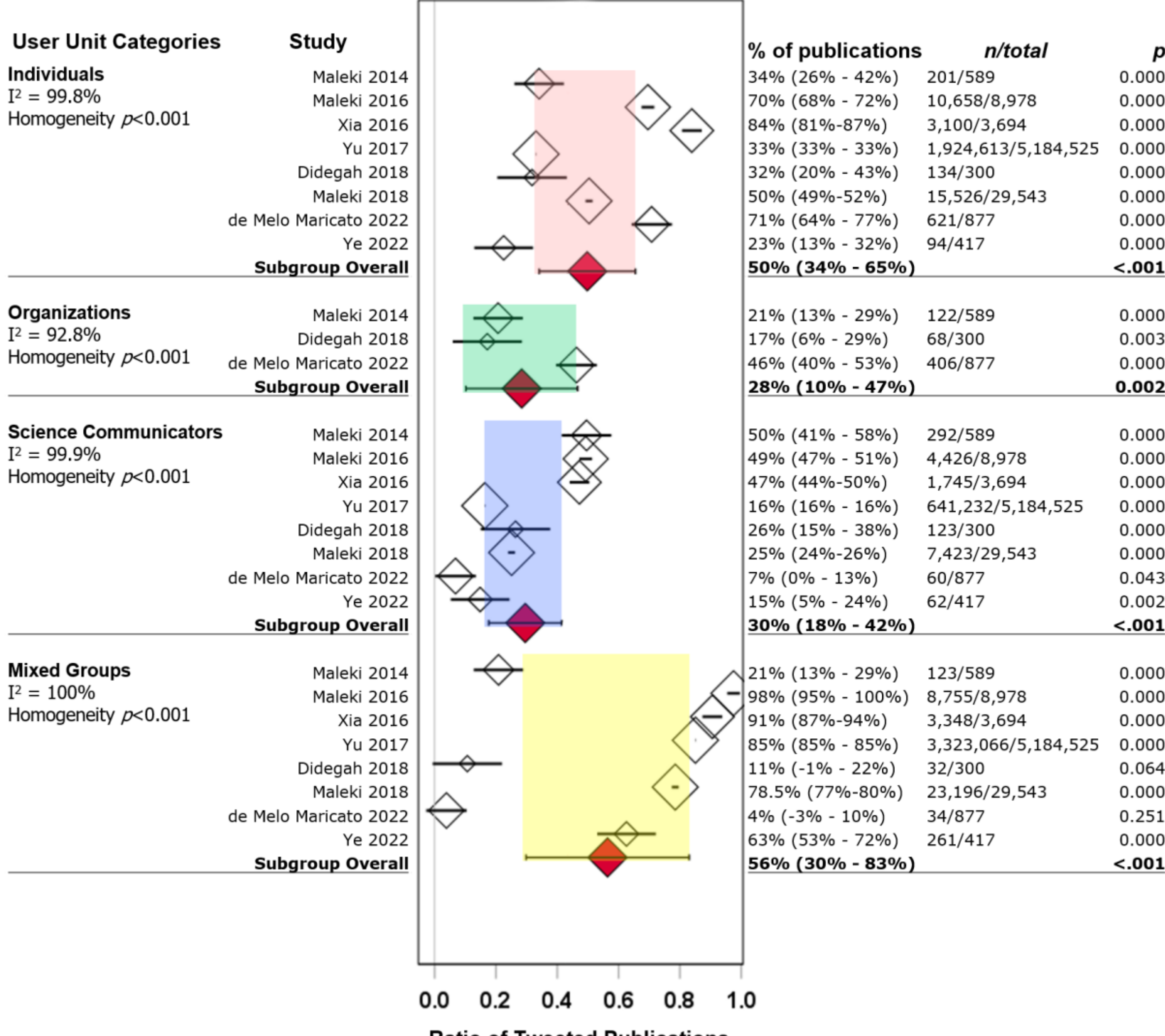

***Figure 8. Meta-analysis of proportion of tweeted publications at User Unit level. The colour ribbons identify the (non-)overlapping studies with the confidence intervals of the overall estimates for each user category.***

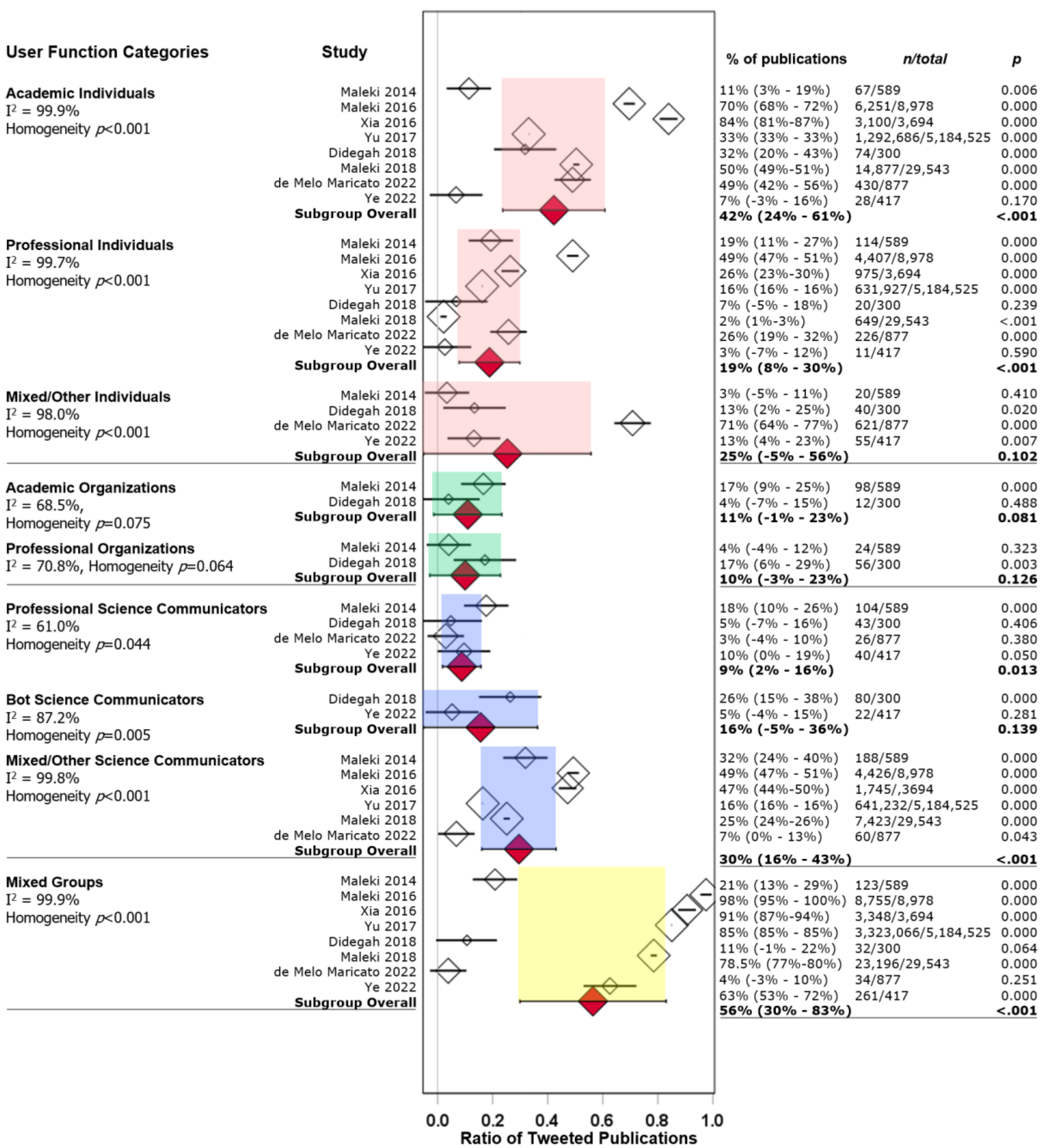

***Figure 9. Meta-analysis of proportion of tweeted publications at User Function level. The colour ribbons identify the (non-)overlapping studies with the confidence intervals of the overall estimates for each user category.***

## 5.5 Synthesis by Beta-Binomial Hierarchical Model (BBHM)

The random-effects meta-analyses above showed substantial heterogeneity ($I^2 > 50\%$ in 40 out of 42 variable-user category combinations), suggesting that study-level differences contribute significantly to the variability in results. Our exploratory meta-analyses included only 23 studies with lower study counts (2-13 studies) per category, making traditional random-effects models susceptible to unreliable variance estimation. Several studies had wide confidence intervals, especially for prorpotion of tweeted publications, with extreme heterogeneity in effect sizes (see significance intervals in forest plots, Figures 4-9).

To address these limitations, we implemented a Bayesian Hierarchical Model (BHM) using a Beta-Binomial structure. This model allows for robust estimation of between-study variability by incorporating hierarchical priors, mitigating overfitting in small-sample meta-analyses, and accommodating classification uncertainty. The model structure estimates category-specific parameters: α (support for users being in-category) and β (support for users being out-of-category), from which the posterior mean engagement proportion ($p$) is derived. The mean value for study variables that were included as binary priors in all the six implemented BBHM models spanned between -0.008 and 0.007 with all HDI intervals overlapping with zero value, suggesting that these study-level characteristics did not exert a consistent or statistically meaningful influence on user engagement estimates across categories, and thus were not strong explanatory factors for between-study variability in the Bayesian framework.

Figure 10 displays both the posterior distributions for α and β and the sampling trace plots for MCMC convergence for selection of *Individuals* in the three examined variables. The left panel illustrates the density of each parameter, while the right panel visualizes draw stability across chains, demonstrating the model's robustness and interpretability for single-category inference (See Supplementary File Figure S1 for posterior distributions of other user categories and variables).

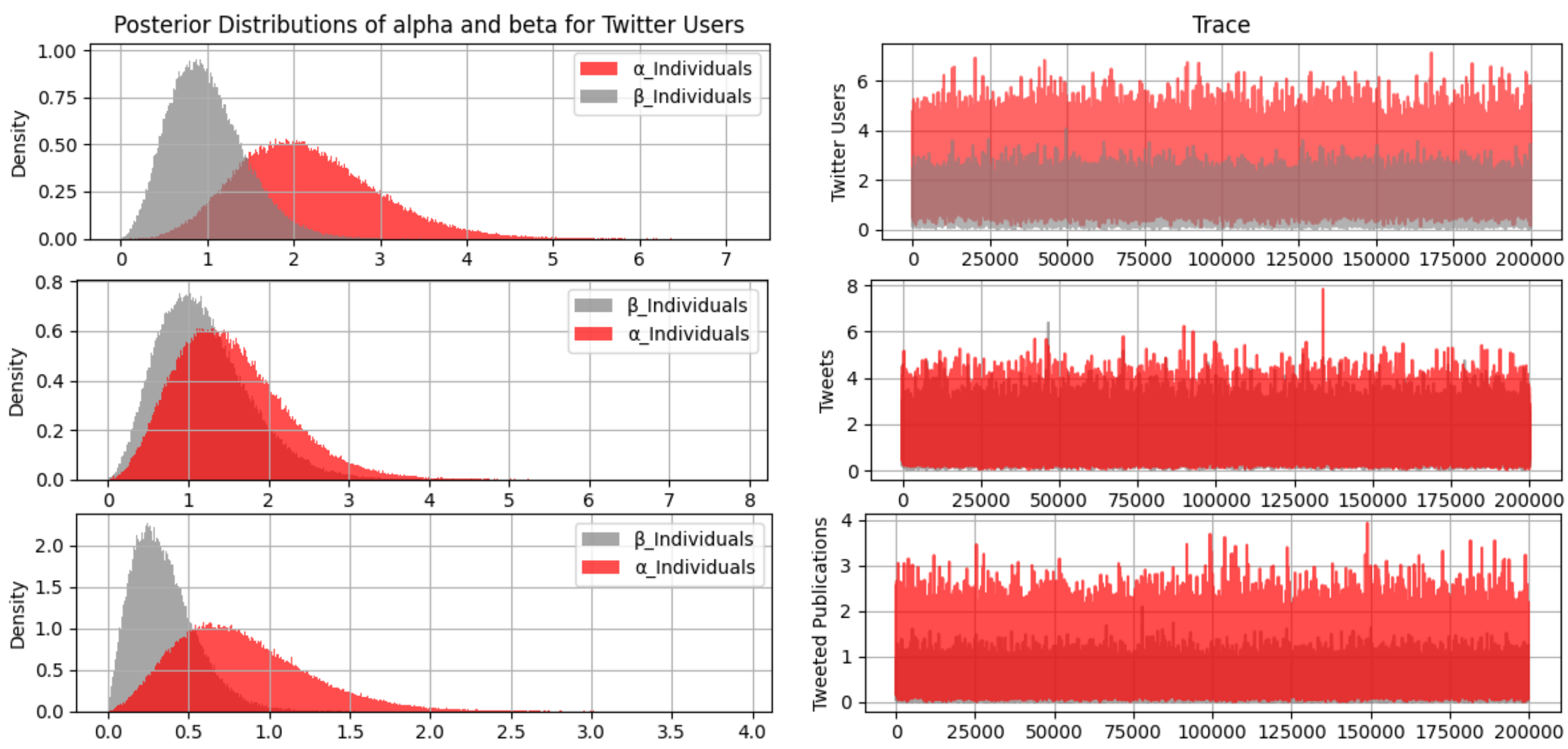

***Figure 10. Posterior Distributions and Sampling Traces for Alpha (in-category probability) and Beta (out-of-category probability) Parameters of Individuals across the three variables of users, tweets and publications from the BBHM analyses. Left: posterior densities for α (red) and β (gray). Right: MCMC sampling trace plots for 50,000 draws across four chains. These plots illustrate parameter convergence and engagement inference within the BBHM framework.***

## 5.5.1 Meta-analysis of Twitter users

Figure 11 shows the BBHM posterior estimates for Twitter user proportions (values detailed in Table S2 in appendix). Among individual user types, academic individuals had an estimated engagement proportion of 33% (α = 1.303, HDI: 0.433–2.224; β = 2.635, HDI: 1.012–4.381). Professional individuals followed with 19% (α = 0.589, β = 2.554), and mixed/unclassified individuals with 24%. "*Individuals excluding Academics*" showed 35%, while the overall Individuals group had the highest engagement at 68% (α = 2.172; β = 1.003).

Organizational user groups contributed far less: academic organizations (10%), professional organizations (19%), and mixed organizations (16%). Among science communicators, academic (12%), bot (8%), and other types (ranging from 9–13%) showed limited engagement.

Mixed groups – unclassified accounts – had engagement proportions of 19% (function level) and 12% (unit level), which shows the prevalence of ambiguous or hybrid user identities across studies.

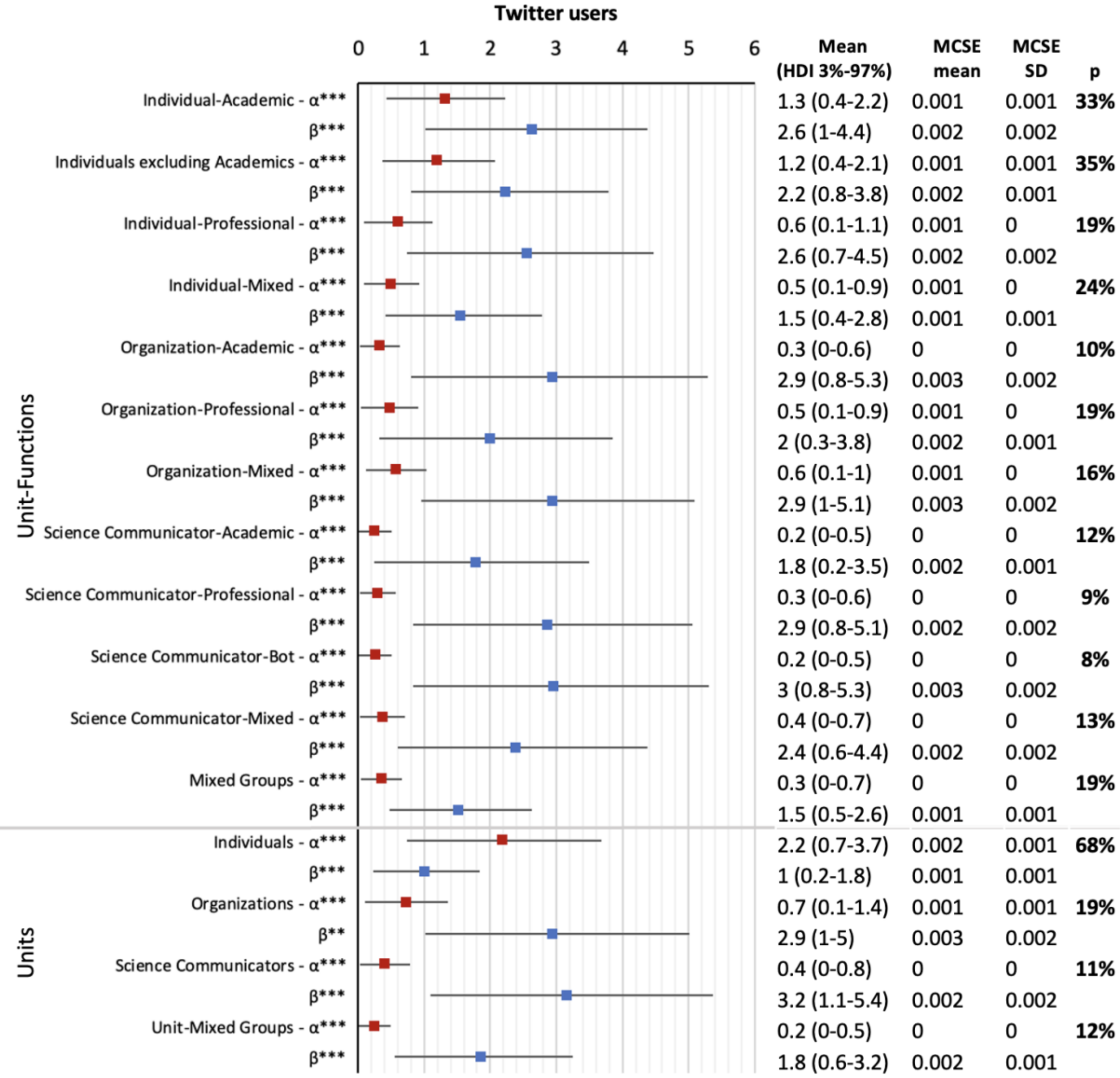

***Figure 11. Posterior mean estimates of Twitter users and 94% highest density intervals (HDI 3%–97%) for the parameters α (red) and β (blue) from two BBHM analyses across unit- and function-level user classification schemes. The right column shows the posterior mean proportion p. Asterisks denote Monte Carlo Standard Error (MCSE) thresholds: *** for MCSE ≤ 0.001 (very high precision), ** for 0.001 < MCSE ≤ 0.005 (high precision). MCSE SD: Standard Deviation of Monte Carlo Standard Error***

### 5.5.2 Meta-analysis of Tweets

Figure 12 presents BBHM results for tweet proportions (values detailed in Table S3 in appendix). *Academic individuals* contributed 22% of tweets ($\alpha = 0.764$; $\beta = 2.639$), *professional individuals* 16%, and *mixed individuals* 30%. "*Individuals excluding Academics*" had 34%, and the broader *Individuals* group reached 56% ($\alpha = 1.529$; $\beta = 1.197$). Organizational user categories remained low: academic organizations (18%), professional organizations (21%), and mixed organizations (15%). Science communicators showed the lowest engagement overall (10%), with bots, mixed, and professional communicators around 12–15%. Mixed groups had a moderate share (31%), with narrower HDIs indicating greater consistency in tweet activity across studies that remain unclassified.

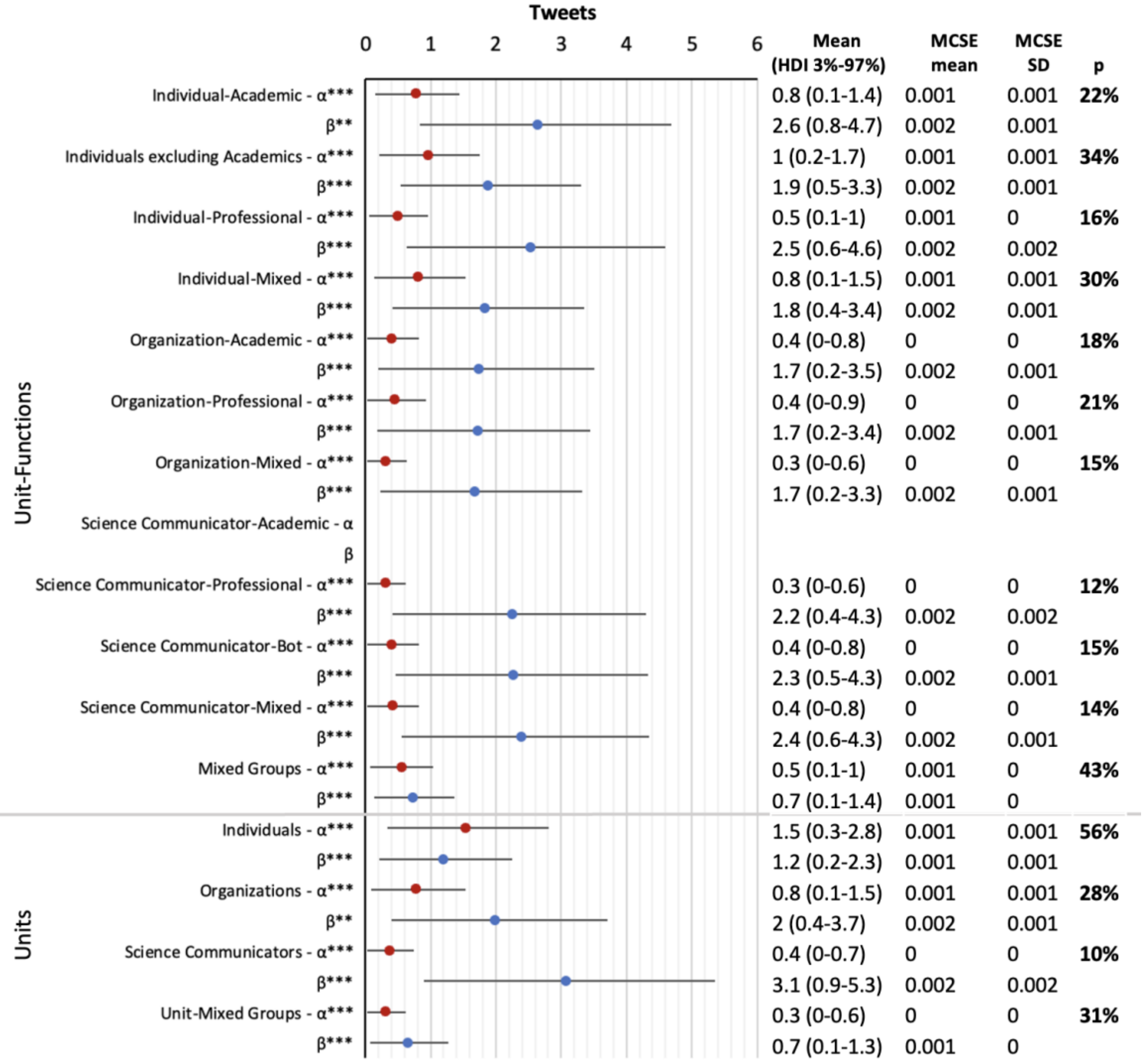

***Figure 12. Posterior mean estimates of Tweets and 94% HDIs for α and β parameters two BBHM analyses across unit- and function-level user classification schemes. See Figure 11 caption for full explanation.***

### 5.5.3 Meta-analysis of Tweeted Publications

Figure 13 summarizes BBHM estimates for tweeted publications (values detailed in Table S4 in appendix). *Academic individuals* had the highest function-specific share (39%; $\alpha = 0.729$, $\beta = 1.157$), followed by *mixed* (28%) and *professional individuals* (17%). "*Individuals excluding Academics*" reached 33%, while all *individuals* combined accounted for 69% of publications shared ($\alpha = 0.856$, $\beta = 0.392$). Organizational groups posted fewer publications: academic (24%) and professional organizations (25%). Science communicators' estimates ranged from 19–30%, with bots and mixed communicators each around 28%. Notably, mixed groups outperformed many classified types, contributing 54% of tweeted publications, suggesting great proportion of publications with ambiguous user engagement.

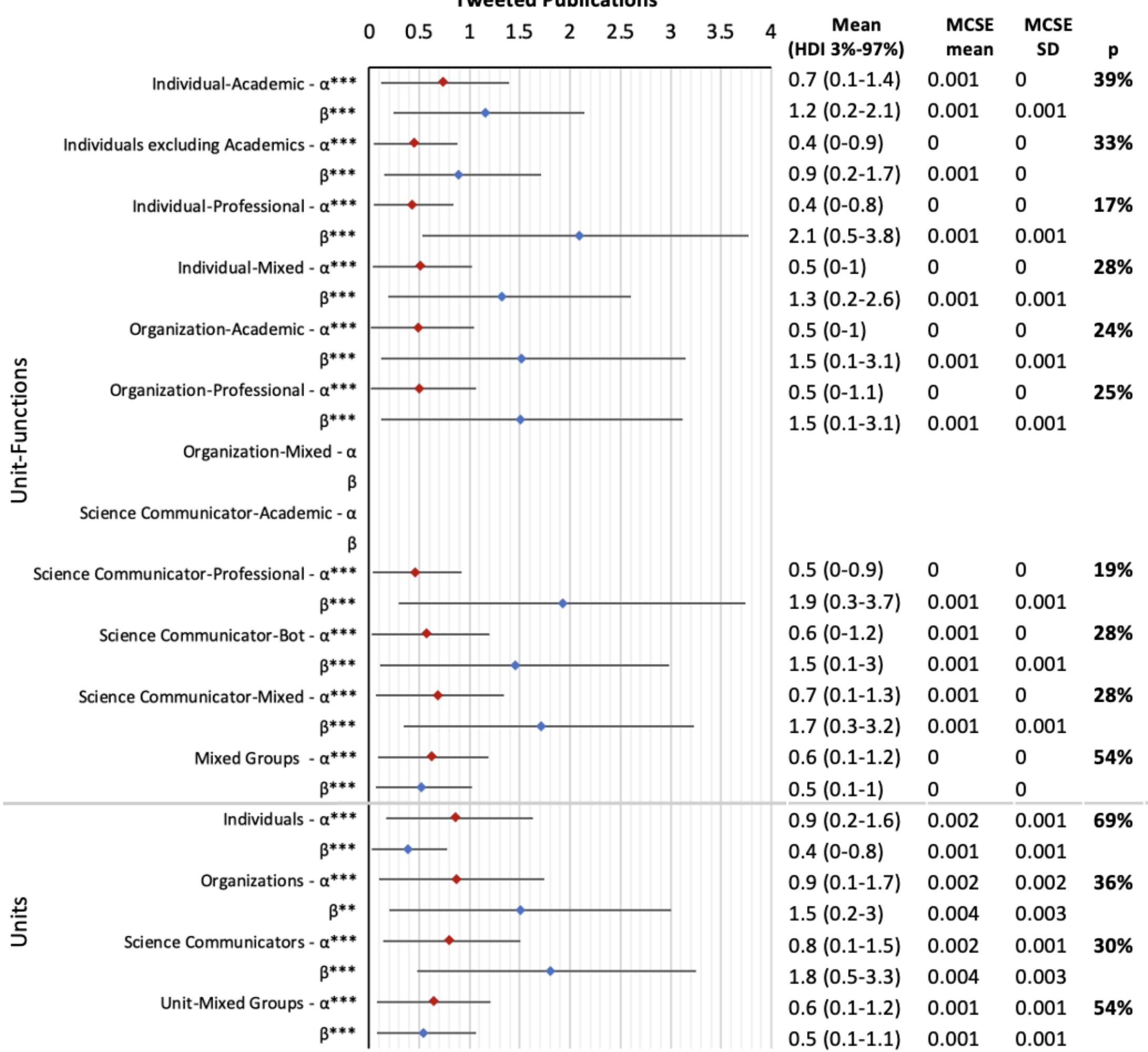

***Figure 13. Posterior mean estimates of Tweeted Publications and 94% HDIs for α and β parameters from two BBHM analyses across unit- and function-level user classification schemes. See Figure 11 caption for full explanation.***

## 5.6 Comparing Results of BBHM and REM Models

As shown in Figure 14, both BBHM and REM identified the same dominant, well-represented contributors – academic individuals and individuals overall – across all engagement variables. For instance, estimates for academic individuals were aligned: 33% (Twitter Users), 22% (Tweets), and 39% (Tweeted Publications). Similarly, individuals overall scored highest in all models: 68% (BBHM) vs. 65% (REM) for users; 56% vs. 55% for tweets; and 69% vs. 50% for tweeted publications.

However, discrepancies between BBHM and REM emerged in groups with sparse data or greater between-study variation. BBHM estimates were typically higher and more stable, while REM often yielded lower proportions and wider confidence intervals, sometimes extending to below-zero values. For academic organizations, BBHM estimated 18% (Tweets) and 24% (Publications), while REM yielded lower values (9% and 11%). For professional organizations: 21% and 25% (BBHM) vs. 12% and 10% (REM).

Science communicator subtypes showed similar patterns. BBHM estimated 12% (Tweets) and 19% (Publications) for professional communicators, compared to 5% and 9% (REM). For bots, BBHM reported 15% and 28%, while REM returned 11% and 16%. These differences highlight REM's tendency to underrepresent smaller or inconsistent groups due to its reliance on between-study variance. BBHM estimates still fall in the estimated intervals in REM although much closer to the higher end.

Mixed user groups were consistently high across both models: 19%, 31%, and 54% (BBHM) vs. 20%, 30%, and 56% (REM). This consistency across models shows the significant role of unclassified user accounts in research dissemination – highlighting the limitations of user classification and rigid typologies in capturing social media engagement.

Overall, BBHM's hierarchical structure and ability to use the strength across studies enabled it to produce plausible, bounded, and stable estimates, even in categories with limited data. In contrast, REM's performance was constrained by its sensitivity to sample size and between-study heterogeneity, leading to lower estimates in smaller groups and less interpretability in confidence intervals. These findings suggest that while REM remains a valid benchmark, BBHM offers reliable modeling for altmetrics research, especially due to complexity of datasets where user types were diverse, fragmented, and classification uncertainty was prevalent.

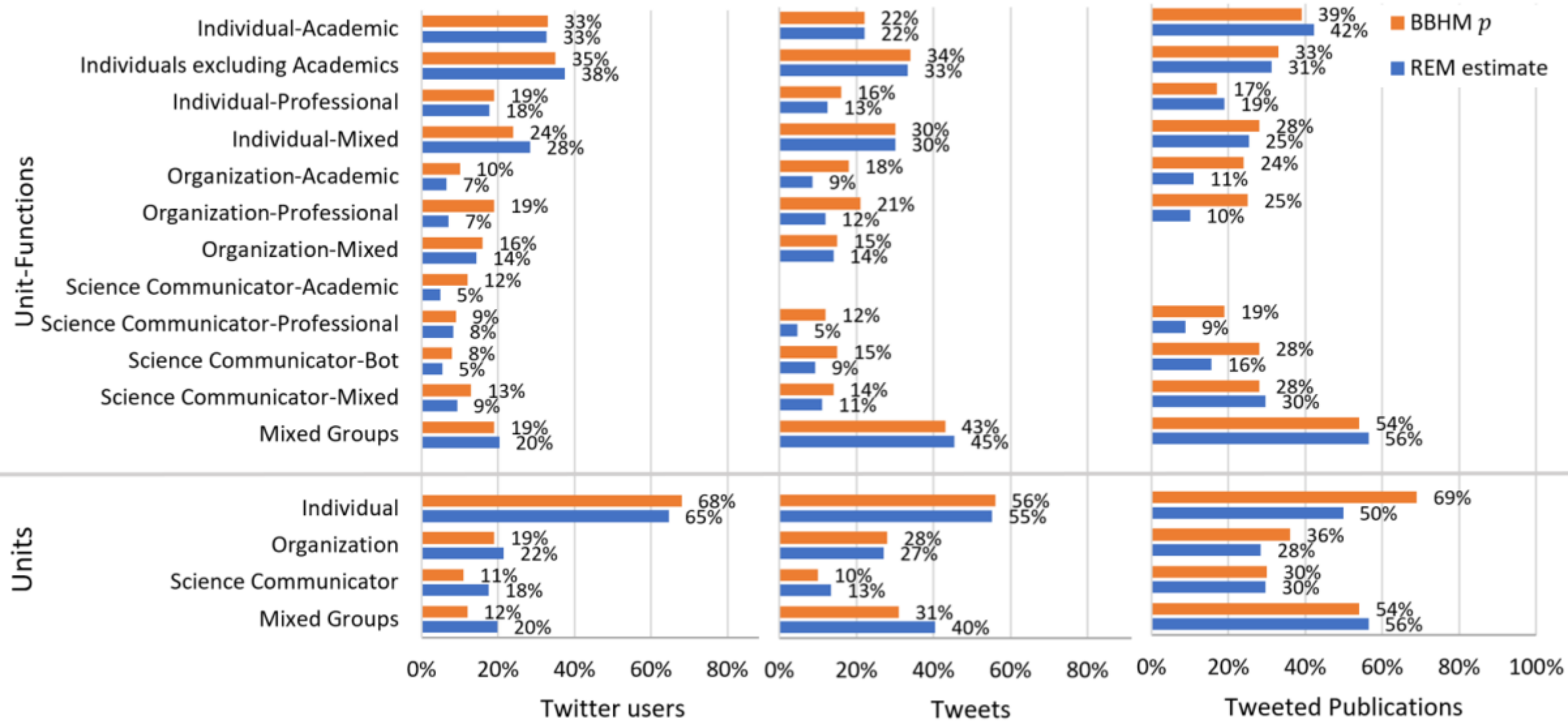

***Figure 14. Comparison between the overall proportion estimates of unit- and function-level user categories in the three variables across REM and BBHM models***

## 5.7 The Mean Difference between User Categories

To further assess which user types are more active in disseminating academic publications compared to academic individuals, a series of mean difference analyses were conducted using study-level proportions. Figure 15 illustrates the mean differences between *Academic Individuals* and all other user categories. In terms of user account proportions, *Academic Individuals* significantly outnumbered *Organizational* accounts by 12% and *Science Communicators* by 15% ($p < 0.01$ for both). However, there was no significant difference between *Academic Individuals* and *Individuals excluding Academics* (I⊄A), indicating that other types of individuals (e.g., professionals or unaffiliated users) are just as present on Twitter as academics themselves.

For tweet activity, *Academic Individuals* tweeted significantly more than *Science Communicators* by 8%, though the confidence interval (8%–18%) indicates that the result was significant only in one-tailed testing ($p < 0.05$). However, they did not tweet significantly more than *Organizational* accounts. Interestingly, *Academic Individuals* posted 12% fewer tweets than *Individuals excluding Academics* (significant from one side with 95% CI: -24% to -12%), indicating that while academics are visible, they may not be the most prolific tweeters.

With regard to tweeted publications, most comparisons lacked sufficient evidence due to limited studies. The only statistically reliable finding was that *Academic Individuals* tweeted 33% more publications than *Professional Science Communicators* ($p < 0.01$), suggesting that researchers are more engaged in direct research dissemination than publishers, media outlets, or journals.

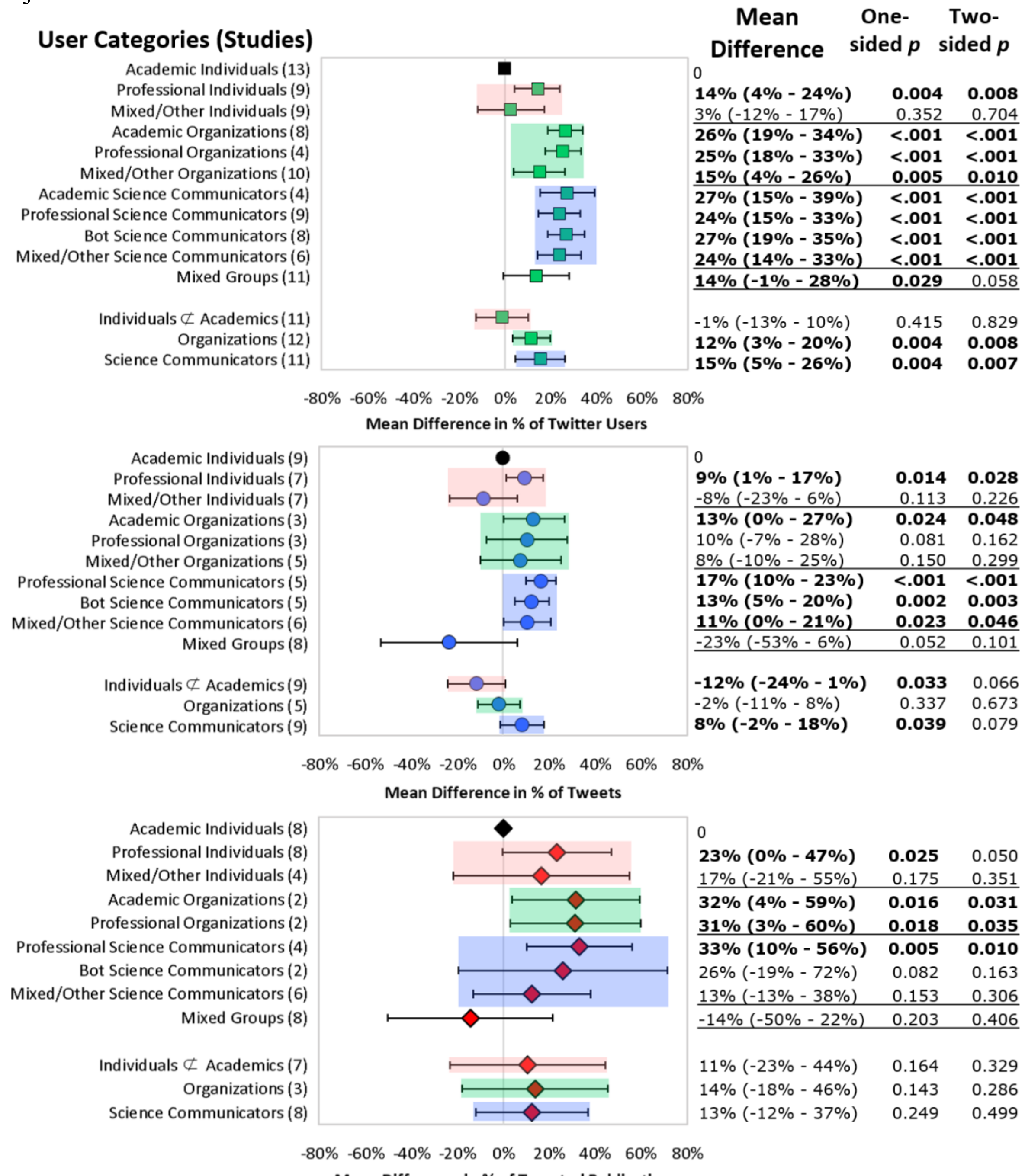

***Figure 15. The mean difference between proportion metrics for Academic Individuals and all the other user categories (both function and unit categories). Significant mean differences are shown in bold. Academic Individuals outnumbered the other categories in positive mean differences and vice versa in negative mean differences. Only when the percentages fall on the side of the mean that error bar has no overlapping with the 0% is significant in one-tailed significant differences. ⊄: not contains. The coloured ribbons indicate significance intervals across user units***

# 6 Limitations

This systematic review and meta-analysis is subject to several limitations, which stem from the nature of the included studies, methodological heterogeneity, and limitations in available data.

A key limitation involves the possibility of overlapping datasets across studies. Due to the lack of access to study-level datasets, we could not determine the extent of duplication in user or publication samples. For instance, Zhou & Na (2019) analyzed 42,891 psychology publications from 2011–2015, which could partially overlap with the 37,627 psychology publications in 2014-2015 examined in Udayakumar, Senadeera et al. (2018), despite differences in sampling strategies and user categorization methods. Similarly, Yu (2017) included data from 27 disciplines between 2011–2016, likely overlapping with multiple studies during the same period.

Another significant limitation lies in the inconsistent and subjective categorization of Twitter users across earlier studies. Some merged individuals and organizations, others used automated vs. manual classification, and hybrid roles (e.g., "practitioner researchers") were often categorized differently. Despite efforts to recategorize users into a unified scheme (unit and function), ambiguities remained – particularly within the Mixed/Other categories. These inconsistencies may have introduced misclassification errors, especially where studies differed in how bots, media, journalist, or science communicators accouts were defined and grouped.

This study consistently observed high statistical heterogeneity ($I^2 > 90\%$) across most user categories and engagement metrics. Although the primary meta-regression revealed some contributing factors – namely the number of user categories, coding method, sample size, and user activity level. For many user groups, insufficient studies limited our ability to fully model and explain the sources of variability.

Several included studies explicitly focused on highly tweeted articles (e.g., Htoo & Jin-Cheon, 2017) or prolific Twitter users (e.g., Haustein et al., 2016; Haustein, 2019; Yu, Xiao et al., 2019). This may have introduced bias, as such users represent only a small proportion of Twitter's user base (Haustein, 2019). The findings indicate that studies centered on high-activity users often reported more bots and higher proportions of unclassified users, suggesting that these may fail to represent the full diversity of Twitter user types. Consequently, findings may be biased toward specific engagement patterns, limiting generalizability to broader Twitter populations.

Initial attempts to use meta-regression to evaluate study-level covariates were limited by small category sizes and failed to produce stable estimates. As a result, binary representations of study-level variables were incorporated into the Beta-Binomial Hierarchical Model (BBHM) instead. While BBHM offers a flexible framework for modeling sparse data and accounting for uncertainty, it still relies on assumptions about beta-binomial distributions and may oversimplify the real-world complexity of social media behavior. Furthermore, estimates for some categories (e.g., bots, professional science communicators) still carry wide highest density intervals (HDIs) due to limited observations.

# 7 Discussion

This study sets out to address a persistent and unresolved question in altmetrics: who is sharing academic research on Twitter, and how does this reflect academic or societal impact? Despite a decade of empirical inquiry, the lack of standardized methodologies and inconsistent user classifications has prevented the field from reaching a conclusive answer. We conducted the first meta-analysis focused specifically on Twitter user categories involved in scholarly communication, synthesizing 23 studies that offered user-level data, and analyzing three distinct engagement metrics – Twitter users, tweets, and tweeted publications – using both Random Effects Models (REM) and a Beta-Binomial Hierarchical Model (BBHM).

*Individuals Dominate, Academics Lead*. The findings show a clear pattern: individual accounts dominate the sharing of academic publications on Twitter. On average, individuals accounted for 66% of all Twitter users, 55% of tweets, and 50% of tweeted publications. Within this group, *academic individuals* consistently emerged as the largest category with some exception in tweet frequency. This pattern remained stable across both REM and BBHM approaches, indicating the robustness of the results. However, our comparisons between academic and non-academic individuals revealed near parity in their contributions: both groups averaged around 30% across most indicators, although academic users showed slightly higher proportions in tweeted publications (~40%) and slightly lower frequency of tweets overal (22%). This result aligns with Zhang et al. (2023), who found similar parity when aggregating Twitter users into academic and societal categories, although their broader and overlapping classification scheme complicates direct comparisons. *Science Communicators and Organizations Play Smaller Roles*. Organizations and science communicators – including journals, publishers, and bots – were significantly less active, typically accounting for less than 30% of any engagement metric. Particularly, professional science communicators (e.g., publishers, media) had the lowest engagement levels across all metrics. *The Mixed Groups Always Have a Role*. A significant proportion of publication were disseminated by the "mixed" or unclassified user groups, that could not be reliably categorized in earlier studies due to study design, and ambiguous or incomplete user data.

*Meta-analysis Models*. Given the high heterogeneity found across studies ($I^2 > 50\%$ in 44 of 46 meta-analyses), the BBHM provided more stable and interpretable estimates than traditional REM models. It proved particularly more interpretable in modeling user categories with sparse data or inconsistent reporting (e.g., science communicators), where REM often yielded low or uninterpretable, negative estimates. By incorporating study-level priors and hierarchical structuring, BBHM allowed us to estimate population-level proportions while accounting for between-study variability. The resulting posterior distributions demonstrated high stability and low Monte Carlo standard errors, affirming the precision of our estimates even in the face of limited or inconsistent primary data.

**Implications for Altmetrics and Scholarly Communication.** Our results raise critical implications for using Twitter as a proxy for societal impact. While individual (non-academic) users are active participants, academic users still comprise a large share of engagement, especially at publication-level. This balance complicates the narrative that Twitter altmetrics significantly capture public attention beyond academia. The data suggests a hybrid model of dissemination, where academics and engaged publics both play key roles, and institutional actors and bots contribute less than previously assumed.

*Dynamics in Platform Use*. Reviewers rightly note that the dynamics of platform use – and the extent of academic versus institutional engagement – may have evolved due to recent shifts in science communication strategies. While our synthesis reflects pre-2021 patterns, future research should revisit these analyses with updated datasets and explore how user engagement patterns evolve across newer platforms (e.g., Bluesky) or institutional outreach strategies. Our findings offer a benchmark, not a ceiling, and emphasize the importance of adaptive classification systems and temporal validation in altmetric research.

# 8 Conclusion

While our results highlight the significance of academics in disseminating research online, and with that, questioning the applicability of Twitter (or X) as a source to investigate societal impact of research, our contributions also have more practical implications for altmetrics research. A major contribution of this study lies in the development and application of a harmonized user categorization scheme, derived from previous studies and refined to include both unit-level (e.g., individual, organization) and function-level (e.g., academic, professional, mixed) categories. This two-level structure enabled the integration of highly heterogeneous datasets and facilitated meaningful category analyses. This harmonized categorization system, built from prior studies, provides a structured foundation for future research and more consistent cross-study comparisons. However, its application requires careful consideration of known limitations. We recommend that future user classification efforts distinguish between unit-level categories (e.g., individuals vs. organizations) and function-level roles (e.g., academics, applied researchers, non-research professionals such as journalists, science communicators including journals, publishers, bots, and aggregators, and other or unclassified users). This two-level scheme helps clarify ambiguities (see Table 6) and offers a scalable approach for improving the precision and interpretability of user-based altmetric analyses.

***Table 6. Suggested Twitter user classification scheme for future studies. The table presents a two-level categorization framework aiming to standardize classification across studies and enable consistent treatment of hybrid roles, across both individuals and organizations.***

| *Unit* / *Function* | *Individuals* | *Organization* |
|---|---|---|
| *Academic* | Academic scientists | university |
| *Non-academic or applied researchers* | Practitioner researchers<br>Industry researchers<br>Political researcher | government organizations, companies, hospitals |
| *Non-research professionals* | All other professionals including<br>Studends and journalists | Non-research organizations<br>Media outlets |
| *Science Communicators* | Aggregators, bloggers, personal or no identity bots | journals, publishers, organizational bots |
| *Others* | The general public not in above roles | |
| *Unrecognizable* | No user information | No user information |

Methodologically, this study introduces the Beta-Binomial Hierarchical Model (BBHM) for altmetric meta-analysis. BBHM proved especially effective in estimating category-specific engagement levels in the presence of data sparsity and heterogeneity. Compared to the Random Effects Model (REM), BBHM generated more stable and interpretable, meaningful results, making it a strong candidate for future meta-analyses in altmetric research. Nevertheless, the study is not without limitations. Overlapping datasets and varying user categorization approaches across studies may have introduced bias. While beta-binomial hierarchical modelling confirmed the stability of our estimates, the structural limitations inherited from earlier studies highlight the need for open data and transparent reporting in future studies.

Our results underscore that design choices – such as selecting tweets, users, or publications as the unit of analysis – can influence interpretations of engagement. For instance, while user-based metrics reflect equal academic and non-academic involvement, tweet-based metrics often highlight more public engagement due to substantially larger scale of public engagement in high profile publications. Thus, future research should carefully consider the metric used when drawing conclusions about impact.

In sum, this study contributes a methodological and conceptual framework for future altmetrics research by demonstrating the value of harmonized user categorization, and robust statistical synthesis and Bayesian modeling. While our findings represent pre-2021 patterns on Twitter (now X), they lay the groundwork for tracking temporal changes and guiding classification practices in emerging platforms such as Bluesky and others. Meta-analysis remains a vital tool for retrospective analysis of online engagement with research, and we encourage future studies to revisit these patterns as the scholarly communication ecosystem continues to evolve.

## 9 Author Contributions

**Ashraf Maleki:** Conceptualization, Data curation, Formal analysis, Investigation, Methodology, Project administration, Resources, Software, Validation, Visualization, Writing – original draft, Writing – review & editing. **Kim Holmberg:** Conceptualization, Funding acquisition, Resources, Supervision, Writing – original draft, Writing – review & editing.

## 10 Competing Interests

The authors have no conflicting interests.

## 11 Funding Information

This study is part of the research project *Applicability of altmetrics in research impact assessment*, funded by the Academy of Finland (332961), currently known as Research Council of Finland.

## 12 Data Availability

All statistics used in the analyses for this systematic review and meta-analysis are fully documented within the main tables presented in the paper. These include user category proportions, study-level outcome measures, and model estimates across both Random Effects Models (REM) and Beta-Binomial Hierarchical Models (BBHM). Should additional clarification or access to structured data be required, for purposes such as replication, reanalysis, or extended exploration, these materials can be made available by the corresponding author upon reasonable request.

# 14 Appendix

## 14.1. Mapping Twitter user categories in studies to our structured recategorization

***Table S1. Study-level Twitter user categories assigned to each category in our structured user categorization***

| Study (Unit / Function) | Individual | | | Organization | | | Science Communicator | | | | Mixed Groups |
|---|---|---|---|---|---|---|---|---|---|---|---|
| | **Academic** | **Professional** | **Other/Mixed** | **Academic** | **Professional** | **Other/Mixed** | **Academic** | **Professional** | **Bot** | **Other/Mixed** | |
| **Maleki 2014** | Official researchers | Professionals and graduates | Public Individuals<br>Students | Science organizations | Product or service providers | | | Media/Media activists<br>Journals/publishers | | Topical article updater | |
| **Tsou 2015** | Ph.D. | | Individuals including students | universities | corporations | Organizations<br>(non-)profit corporations | | news/media/outreach | | | unknown |
| **Haustein Tsou 2016** | Researcher | Professional | Individual<br>Student | Scientific Society/Association<br>College/University<br>Research Center | Corporation | Non-profit organizations<br>Government<br>K-12<br>Unable to tell<br>Other | Library | Journal Publisher | Completely automated bots | Science communicator<br>Partially automated bots | |
| **Haustein Bowman 2016** | | | | | | | | | self-identified bots<br>Platform feed | Topic feed<br>Selective/qualitative | |
| **Maleki 2016** | Scientist | Practitioner | | | | | | | | Science communicator | member of the public |
| **Xia 2016** | Scientist | Practitioner | | | | | | | | Science communicator | member of the public |
| **Htoo 2017** | Academic individual | | Non-academic individual | Academic organization | | non-academic organization | | | | | |
| **Vainio 2017** | Researcher<br>Post-doc/PhD<br>Professor | Health care professional<br>Entrepreneur<br>Position of expertise<br>Other profession | student | Government organization/Universities | non-profit organization/non-profit group<br>Company | Opinion/Propaganda | Librarian/other academic | Writer/editor/journalist<br>Publisher | | Publication (not peer-reviewed) | Other |
| **Yu 2017** | Scientist | Practitioner | | | | | | | | Science communicator | member of the public |
| **Didegah 2018** | Individual researcher | Individual professionals | Individual citizen | Research organization | Business<br>Civil society organization<br>Intergovernmental organization | Research Funding | | Individual journalist<br>Media<br>Publishers/journals | Bots | Mixed (bot and non-bot) | Other or blank non-bot |
| **Maleki 2018** | Scientist | Practitioner | | | | | | | | Science communicator | member of the public |
| **Sergiadis 2018** | affiliated colleagues<br>affiliated authors | | unaffiliated individuals | affiliated organizations | | unaffiliate organizations | | | | | |
| **Udayakumar 2018** | Academic | non-Academic | | Academic Organization | non-Academic organization | Organization | | | | | |
| **Haustein 2019** | | | | | | | | Journal and publisher accounts | Self-identified bots | | Other accounts |
| **Toupin 2019** | Faculty members and students<br>Personal-Faculty and Students | Professionals | Personal keywords unique (non-overlapping part) | | | Institutions and organizations | | Journals and publishers<br>Communicators and journalists | Bots and automated accounts | | Unidentified users |

| Study | Unit | Individual | | | Organization | | | Science Communicator | | | | Mixed Groups |
|---|---|---|---|---|---|---|---|---|---|---|---|---|
| | Function | Academic | Professional | Other/Mixed | Academic | Professional | Other/Mixed | Academic | Professional | Bot | Other/Mixed | |
| **Yu 2019** | | Researcher<br>1 University faculty<br>2 PhD student<br>3 Post-doc researchers<br>4 Researcher in scientific institution<br>5 Researcher in enterprises<br>6 Other types of institutional researcher<br>7 Other types of individual researcher | 12 Non-researchers from other types of institutions | 11 Non-researcher individuals | 8 Universities (Institutional account)<br>9 Scientific institution (Institutional account) | 10 R&D enterprises (Institutional account)<br>17 Other types of institutions (Institutional account) | Institutional account | 13 Librarian<br>15 Libraries (Institutional account) | 14 Journals (Institutional account)<br>16 Publishers (Institutional account) | Bot account | Cyborg account<br>Science Communicator | |
| **Zhou 2019** | | Academic individuals | | Non-Academic individuals | Academic organization | | Non-academic organizations | | | | | Unknown |
| **Haustein 2020** | | Researcher<br>Professor/Lecturer | Business/Professional | Student<br>Not enough information | University/ Research Centre | Database/Platform | Association/ Organization | Library/Librarian | Outreach Journalist<br>Journal Publisher | Bot | | Not enough information |
| **Lemke 2021** | | | | Individual | | | Group | | | | | Unidentifiable |
| **Orduña-Malea 2021** | | | | individual | | | Institutional | | | | | Undefined |
| **Abhari 2022** | | Scientists | Practitioners | Public | | | | | | Bots | Science Communicators | remaining |
| **de Melo Maricato 2022** | | Researcher (Including Postdoctoral)<br>Professor<br>Biologist<br>Student (PhD, MSc and others)<br>Ecologist<br>Lecturer<br>Geographer and Geologist<br>Economist<br>Microbiologist<br>Agronomist<br>Historian | Doctor<br>Arts (Writer etc.)<br>Psychologist<br>Manager<br>Physician<br>Physiotherapist<br>Lawyer<br>Academic manager<br>Business(wo)man<br>Coach<br>Sports professions<br>Consultant<br>Veterinarian<br>Executive<br>Chemical<br>Information Technology<br>Teacher<br>Educator<br>Engineer<br>Nutritionist<br>Dentist<br>Pharmacist | Personal | | | Organizational | | Journalist<br>Science communicator<br>Journal Editor | | Character | No Information |
| **Ye 2022** | | Academic researchers & institutions – Bot | Health science practitioners<br>Non-health science practitioners | | | | | | Academic publishers<br>Mass media | Bots | Research feeds<br>Topic feeds & news alerts | Other |

# 14.2. Beta-Binomial Hierarchical Model Posterior Distributions

***Table S2. Summary of Bayesian Hierarchical Model Estimates for Percent of Twitter Users at unit and function Levels.***

| Twitter Users | | mean | sd | hdi_3% | hdi_97% | mcse_mean | mcse_sd | ess_bulk | ess_tail | r_hat | $p$ |
|---|---|---|---|---|---|---|---|---|---|---|---|
| **Units** | | | | | | | | | | | |
| Individuals | α | 2.172 | 0.805 | 0.743 | 3.683 | 0.002 | 0.001 | 146,527 | 109,990 | 1 | 0.684 |
| | β | 1.003 | 0.444 | 0.231 | 1.832 | 0.001 | 0.001 | 134,692 | 101,578 | 1 | |
| Organizations | α | 0.711 | 0.352 | 0.107 | 1.353 | 0.001 | 0.001 | 122,077 | 91,569 | 1 | 0.195 |
| | β | 2.937 | 1.098 | 1.024 | 5.013 | 0.003 | 0.002 | 147,014 | 110,358 | 1 | |
| Science Communicators | α | 0.388 | 0.221 | 0.031 | 0.789 | 0 | 0 | 163,956 | 98,665 | 1 | 0.109 |
| | β | 3.157 | 1.171 | 1.101 | 5.363 | 0.002 | 0.002 | 216,346 | 123,130 | 1 | |
| Unit-Mixed Groups | α | 0.241 | 0.144 | 0.013 | 0.5 | 0 | 0 | 156,978 | 91,455 | 1 | 0.115 |
| | β | 1.849 | 0.741 | 0.557 | 3.245 | 0.002 | 0.001 | 182,308 | 106,918 | 1 | |
| **Unit-Function** | | | | | | | | | | | |
| Individual-Academic | α | 1.303 | 0.493 | 0.433 | 2.224 | 0.001 | 0.001 | 133,573 | 110,820 | 1 | 0.331 |
| | β | 2.635 | 0.926 | 1.012 | 4.381 | 0.002 | 0.002 | 139,068 | 116,979 | 1 | |
| Individuals excluding Academics | α | 1.185 | 0.467 | 0.372 | 2.065 | 0.001 | 0.001 | 128,885 | 107,534 | 1 | 0.347 |
| | β | 2.232 | 0.819 | 0.798 | 3.785 | 0.002 | 0.001 | 140,133 | 117,150 | 1 | |
| Individual-Professional | α | 0.589 | 0.291 | 0.095 | 1.126 | 0.001 | 0 | 159,287 | 99,947 | 1 | 0.187 |
| | β | 2.554 | 1.042 | 0.734 | 4.476 | 0.002 | 0.002 | 189470 | 118,544 | 1 | |
| Individual-Mixed | α | 0.487 | 0.234 | 0.092 | 0.922 | 0.001 | 0 | 159,589 | 102,976 | 1 | 0.240 |
| | β | 1.546 | 0.661 | 0.415 | 2.78 | 0.001 | 0.001 | 173,131 | 114,592 | 1 | |
| Organization-Academic | α | 0.323 | 0.171 | 0.036 | 0.635 | 0 | 0 | 151,024 | 91,808 | 1 | 0.099 |
| | β | 2.945 | 1.258 | 0.8 | 5.294 | 0.003 | 0.002 | 203,776 | 121,351 | 1 | |
| Organization-Professional | α | 0.463 | 0.25 | 0.05 | 0.917 | 0.001 | 0 | 179,521 | 102,644 | 1 | 0.189 |
| | β | 1.991 | 1.016 | 0.326 | 3.848 | 0.002 | 0.001 | 225,878 | 119,129 | 1 | |
| Organization-Mixed | α | 0.561 | 0.255 | 0.124 | 1.035 | 0.001 | 0 | 131,441 | 93,515 | 1 | 0.160 |
| | β | 2.94 | 1.148 | 0.952 | 5.09 | 0.003 | 0.002 | 167,935 | 120,841 | 1 | |
| Science Communicator-Academic | α | 0.243 | 0.144 | 0.017 | 0.503 | 0 | 0 | 185,760 | 98,019 | 1 | 0.120 |
| | β | 1.781 | 0.943 | 0.248 | 3.499 | 0.002 | 0.001 | 208,913 | 111,716 | 1 | |
| Science Communicator-Professional | α | 0.288 | 0.16 | 0.026 | 0.577 | 0 | 0 | 157,243 | 94,896 | 1 | 0.091 |
| | β | 2.867 | 1.174 | 0.826 | 5.053 | 0.002 | 0.002 | 238,902 | 123,876 | 1 | |
| Science Communicator-Bot | α | 0.246 | 0.146 | 0.015 | 0.508 | 0 | 0 | 177,126 | 101,196 | 1 | 0.077 |
| | β | 2.95 | 1.25 | 0.829 | 5.305 | 0.003 | 0.002 | 210,213 | 116,345 | 1 | |
| Science Communicator-Mixed | α | 0.36 | 0.195 | 0.036 | 0.714 | 0 | 0 | 176,654 | 100,513 | 1 | 0.131 |
| | β | 2.387 | 1.063 | 0.598 | 4.373 | 0.002 | 0.002 | 201,676 | 116,700 | 1 | |
| Mixed Groups | α | 0.344 | 0.178 | 0.044 | 0.671 | 0 | 0 | 136,419 | 91,441 | 1 | 0.185 |
| | β | 1.513 | 0.594 | 0.473 | 2.624 | 0.001 | 0.001 | 180,478 | 113,491 | 1 | |

$p$ : Estimated Engagement Proportion ($p = \alpha / (\alpha+\beta)$)

***Table S3. Summary of Bayesian Hierarchical Model Estimates for Percent of Tweets at unit and function Levels.***

| Tweets | | mean | sd | hdi_3% | hdi_97% | mcse_mean | mcse_sd | ess_bulk | ess_tail | r_hat | *p* |
|---|---|---|---|---|---|---|---|---|---|---|---|
| **Units** | | | | | | | | | | | |
| Individuals | α | 1.529 | 0.691 | 0.341 | 2.808 | 0.001 | 0.001 | 189,941 | 118,796 | 1 | 0.561 |
| | β | 1.197 | 0.571 | 0.223 | 2.253 | 0.001 | 0.001 | 176,132 | 109,373 | 1 | |
| Organizations | α | 0.771 | 0.416 | 0.084 | 1.529 | 0.001 | 0.001 | 195,456 | 110,558 | 1 | 0.280 |
| | β | 1.979 | 0.932 | 0.4 | 3.707 | 0.002 | 0.001 | 212,134 | 118,977 | 1 | |
| Science Communicators | α | 0.359 | 0.209 | 0.027 | 0.739 | 0 | 0 | 188,429 | 102,840 | 1 | 0.105 |
| | β | 3.074 | 1.228 | 0.902 | 5.349 | 0.002 | 0.002 | 278,804 | 125,616 | 1 | |
| Unit-Mixed Groups | α | 0.299 | 0.175 | 0.018 | 0.612 | 0 | 0 | 196,671 | 96,714 | 1 | 0.313 |
| | β | 0.656 | 0.341 | 0.082 | 1.276 | 0.001 | 0 | 223,885 | 104,974 | 1 | |
| **Unit-Function** | | | | | | | | | | | |
| Individual-Academic | α | 0.764 | 0.362 | 0.146 | 1.436 | 0.001 | 0.001 | 197,907 | 115,356 | 1 | 0.225 |
| | β | 2.639 | 1.069 | 0.837 | 4.688 | 0.002 | 0.001 | 241,020 | 132,153 | 1 | |
| Individuals excluding Academics | α | 0.953 | 0.43 | 0.211 | 1.749 | 0.001 | 0.001 | 200,480 | 118,842 | 1 | 0.337 |
| | β | 1.871 | 0.767 | 0.547 | 3.312 | 0.002 | 0.001 | 219,183 | 124,658 | 1 | |
| Individual-Professional | α | 0.49 | 0.256 | 0.061 | 0.956 | 0.001 | 0 | 213,727 | 115,157 | 1 | 0.162 |
| | β | 2.535 | 1.113 | 0.632 | 4.602 | 0.002 | 0.002 | 262,326 | 126,145 | 1 | |
| Individual-Mixed | α | 0.801 | 0.397 | 0.139 | 1.542 | 0.001 | 0.001 | 206,986 | 119,388 | 1 | 0.305 |
| | β | 1.829 | 0.821 | 0.425 | 3.357 | 0.002 | 0.001 | 230,611 | 123,259 | 1 | |
| Organization-Academic | α | 0.389 | 0.239 | 0.026 | 0.825 | 0 | 0 | 206,547 | 104,661 | 1 | 0.182 |
| | β | 1.744 | 0.965 | 0.198 | 3.508 | 0.002 | 0.001 | 238,815 | 110,156 | 1 | |
| Organization-Professional | α | 0.446 | 0.269 | 0.029 | 0.934 | 0 | 0 | 230,943 | 108,028 | 1 | 0.206 |
| | β | 1.715 | 0.947 | 0.179 | 3.44 | 0.002 | 0.001 | 275,246 | 117,902 | 1 | |
| Organization-Mixed | α | 0.302 | 0.184 | 0.019 | 0.635 | 0 | 0 | 234,745 | 111,711 | 1 | 0.152 |
| | β | 1.683 | 0.892 | 0.234 | 3.325 | 0.002 | 0.001 | 234,670 | 107,575 | 1 | |
| Science Communicator-Academic | α | 0.3 | 0.176 | 0.019 | 0.617 | 0 | 0 | 240,460 | 108,308 | 1 | 0.118 |
| | β | 2.249 | 1.107 | 0.412 | 4.294 | 0.002 | 0.002 | 262,628 | 119,948 | 1 | |
| Science Communicator-Professional | α | 0.399 | 0.233 | 0.029 | 0.818 | 0 | 0 | 221,322 | 110,621 | 1 | 0.150 |
| | β | 2.268 | 1.095 | 0.463 | 4.323 | 0.002 | 0.001 | 265,051 | 118,883 | 1 | |
| Science Communicator-Bot | α | 0.405 | 0.226 | 0.037 | 0.815 | 0 | 0 | 197,014 | 105,429 | 1 | 0.145 |
| | β | 2.393 | 1.059 | 0.557 | 4.344 | 0.002 | 0.001 | 246,669 | 121,156 | 1 | |
| Science Communicator-Mixed | α | 0.544 | 0.271 | 0.082 | 1.039 | 0.001 | 0 | 233,703 | 113,768 | 1 | 0.430 |
| | β | 0.722 | 0.343 | 0.136 | 1.359 | 0.001 | 0 | 229,817 | 112,263 | 1 | 0.225 |
| Mixed Groups | α | 0.764 | 0.362 | 0.146 | 1.436 | 0.001 | 0.001 | 197,907 | 115,356 | 1 | |
| | β | 2.639 | 1.069 | 0.837 | 4.688 | 0.002 | 0.001 | 241,020 | 132,153 | 1 | 0.337 |

*p* : Estimated Engagement Proportion ($p = \alpha / (\alpha+\beta)$)

***Table S4. Summary of Bayesian Hierarchical Model Estimates for Percent of Tweeted Publications at unit and function Levels.***

| Tweeted Publications | | mean | sd | hdi_3% | hdi_97% | mcse_mean | mcse_sd | ess_bulk | ess_tail | r_hat | ***p*** |
|---|---|---|---|---|---|---|---|---|---|---|---|
| **Units** | | | | | | | | | | | |
| Individuals | α | 0.856 | 0.408 | 0.176 | 1.625 | 0.002 | 0.001 | 39,522 | 21,209 | 1 | 0.686 |
| | β | 0.392 | 0.217 | 0.034 | 0.782 | 0.001 | 0.001 | 32,559 | 20,663 | 1 | |
| Organizations | α | 0.865 | 0.475 | 0.101 | 1.743 | 0.002 | 0.002 | 39,139 | 20,173 | 1 | 0.365 |
| | β | 1.506 | 0.809 | 0.204 | 3 | 0.004 | 0.003 | 42,365 | 21,990 | 1 | |
| Science Communicators | α | 0.789 | 0.38 | 0.146 | 1.505 | 0.002 | 0.001 | 26,848 | 17,988 | 1 | 0.304 |
| | β | 1.806 | 0.776 | 0.477 | 3.256 | 0.004 | 0.003 | 34,407 | 23,598 | 1 | |
| Unit-Mixed Groups | α | 0.637 | 0.319 | 0.077 | 1.21 | 0.001 | 0.001 | 38,317 | 21,212 | 1 | 0.538 |
| | β | 0.546 | 0.279 | 0.083 | 1.069 | 0.001 | 0.001 | 32,672 | 18,793 | 1 | |
| **Unit-Function** | | | | | | | | | | | |
| Individual-Academic | α | 0.729 | 0.36 | 0.119 | 1.392 | 0.001 | 0 | 328,995 | 213,349 | 1 | 0.387 |
| | β | 1.157 | 0.532 | 0.246 | 2.148 | 0.001 | 0.001 | 341,462 | 212,037 | 1 | |
| Individuals excluding Academics | α | 0.444 | 0.24 | 0.046 | 0.879 | 0 | 0 | 368,069 | 214,534 | 1 | 0.333 |
| | β | 0.891 | 0.44 | 0.157 | 1.71 | 0.001 | 0 | 420,108 | 221,403 | 1 | |
| Individual-Professional | α | 0.425 | 0.229 | 0.048 | 0.843 | 0 | 0 | 352,195 | 212,423 | 1 | 0.169 |
| | β | 2.096 | 0.906 | 0.528 | 3.775 | 0.001 | 0.001 | 398,661 | 226,395 | 1 | |
| Individual-Mixed | α | 0.506 | 0.289 | 0.043 | 1.028 | 0 | 0 | 350,577 | 205,464 | 1 | 0.276 |
| | β | 1.325 | 0.695 | 0.193 | 2.607 | 0.001 | 0.001 | 404,249 | 222,746 | 1 | |
| Organization-Academic | α | 0.489 | 0.31 | 0.022 | 1.047 | 0 | 0 | 400,343 | 210,722 | 1 | 0.243 |
| | β | 1.52 | 0.894 | 0.126 | 3.148 | 0.001 | 0.001 | 428,669 | 210,237 | 1 | |
| Organization-Professional | α | 0.499 | 0.315 | 0.023 | 1.065 | 0 | 0 | 413,093 | 220,816 | 1 | 0.249 |
| | β | 1.505 | 0.889 | 0.117 | 3.121 | 0.001 | 0.001 | 420,005 | 214,693 | 1 | |
| Organization-Mixed | α | 0.452 | 0.26 | 0.036 | 0.924 | 0 | 0 | 381,094 | 212,799 | 1 | 0.190 |
| | β | 1.93 | 0.988 | 0.299 | 3.739 | 0.001 | 0.001 | 422,779 | 222,416 | 1 | |
| Science Communicator-Academic | α | 0.569 | 0.351 | 0.031 | 1.2 | 0.001 | 0 | 379,783 | 209,601 | 1 | 0.282 |
| | β | 1.451 | 0.848 | 0.113 | 2.981 | 0.001 | 0.001 | 441,899 | 219,726 | 1 | |
| Science Communicator-Professional | α | 0.677 | 0.367 | 0.07 | 1.346 | 0.001 | 0 | 345,190 | 207,018 | 1 | 0.283 |
| | β | 1.714 | 0.815 | 0.343 | 3.231 | 0.001 | 0.001 | 393,638 | 225,466 | 1 | |
| Science Communicator-Bot | α | 0.616 | 0.312 | 0.089 | 1.188 | 0 | 0 | 398,578 | 213,832 | 1 | 0.539 |
| | β | 0.526 | 0.273 | 0.068 | 1.024 | 0 | 0 | 391,371 | 214,250 | 1 | |
| Science Communicator-Mixed | α | 0.729 | 0.36 | 0.119 | 1.392 | 0.001 | 0 | 328,995 | 213,349 | 1 | 0.387 |
| | β | 1.157 | 0.532 | 0.246 | 2.148 | 0.001 | 0.001 | 341,462 | 212,037 | 1 | |
| Mixed Groups | α | 0.444 | 0.24 | 0.046 | 0.879 | 0 | 0 | 368,069 | 214,534 | 1 | 0.333 |
| | β | 0.891 | 0.44 | 0.157 | 1.71 | 0.001 | 0 | 420,108 | 221,403 | 1 | |

***p*** : Estimated Engagement Proportion (***p*** = α / (α+β))